\documentclass[journal]{IEEEtran} 
 \IEEEoverridecommandlockouts                  
\usepackage[mode=buildmissing]{standalone} 

\usepackage[tbtags]{amsmath}
\usepackage{nccmath}
\usepackage{amssymb}
\usepackage{amsthm}
\usepackage{bm}
\usepackage{color}
\usepackage{verbatim}
\usepackage{balance}
\usepackage{graphicx}

\usepackage{booktabs,siunitx, multirow,ragged2e}
\usepackage{hyperref}
\usepackage{dsfont}
\usepackage{cite}
\usepackage{float}
\usepackage{algorithm}
\usepackage[noend]{algpseudocode}
\usepackage{tikz,pgfplots}
\usepgfplotslibrary{fillbetween}
\usepgfplotslibrary{groupplots}
\makeatletter
\def\BState{\State\hskip-\ALG@thistlm}
\makeatother
\usepackage{xcolor}

\graphicspath{{../figures/}}
\pgfplotsset{compat=newest}%
\usetikzlibrary{positioning,matrix,shapes.multipart,shapes.misc,spy}%
\tikzstyle{annotation}=[fill=white]%
\usepackage{balance}
\usepackage{acro}

\usepackage{xcolor}

\definecolor{VAE_color}{rgb}{0, 1, 0.3}
\definecolor{CMA_color}{rgb}{0,0,0}

\definecolor{MMSE_color}{rgb}{0, 0, 1}

\definecolor{VQ_VAE_color1}{rgb}{1, 0, 0} 
\definecolor{VQ_VAE_color2}{rgb}{1, 0.2, 0.2}
\definecolor{VQ_VAE_color3}{rgb}{1, 0.5, 0.3}

\definecolor{VQ_VAE_color4}{rgb}{1, 0.3, 0.6}

\newcommand{\vect}[1]{\bm{#1}}
\newcommand{\mat}[1]{\bm{#1}}
\newcommand{\transpose}{\top}
\newcommand{\Ctranspose}{\dagger}

\usetikzlibrary{shapes.arrows}
\usetikzlibrary{spy}

\tikzset{%
	partial ellipse/.style args={#1:#2:#3}{%
		insert path={+ (#1:#3) arc (#1:#2:#3)}%
	}%
}%

\DeclareAcronym{AE}{
short=AE,
long= autoencoder,
}

\DeclareAcronym{WDM}{
short=WDM,
long=wavelength division multiplexing ,
}

\DeclareAcronym{NN}{
short=NN,
long= neural network,
}
\DeclareAcronym{DSP}{
short=DSP,
long= digital signal processing,
}
\DeclareAcronym{CMA}{
short=CMA,
long= constant modulus algorithm,
}

\DeclareAcronym{MMA}{
short=MMA,
long= multi-modulus algorithm,
}

\DeclareAcronym{PS}{
short= PS,
long= pulse-shaping,
}

\DeclareAcronym{DAC}{
short=DAC,
long= digital-to-analog converter,
}

\DeclareAcronym{ADC}{
short=ADC,
long= analog-to-digital converter,
}

\DeclareAcronym{DL}{
short=DL,
long= direct learning,
}

\DeclareAcronym{ENOB}{
short=ENOB,
long= effective number of bits,
}

\DeclareAcronym{DPD}{
short= DPD,
long= digital pre-distortion,
}

\DeclareAcronym{}{
short=,
long= ,
}

\DeclareAcronym{MZM}{
short= MZM,
long= Mach-Zehnder modulator ,
}

\DeclareAcronym{MZI}{
short= MZM,
long= Mach-Zehnder interferometer ,
}
\DeclareAcronym{ML}{
short= ML,
long= maximum-likelihood  ,
}
\DeclareAcronym{VAE}{
short= VAE,
long= variational autoencoder  ,
}

\DeclareAcronym{VQ-VAE}{
short= VQ-VAE,
long= vector quantized VAE  ,
}

\DeclareAcronym{EDFA}{
short= EDFA,
long= erbium-doped fiber amplifier ,
}

\DeclareAcronym{SE}{
short= SE,
long= spectral efficiency ,
}

\DeclareAcronym{PA}{
short= PA,
long= power amplifier ,
}

\DeclareAcronym{MF}{
short= MF,
long= matched filter ,
} 

\DeclareAcronym{OOB}{
short= OOB,
long= out-of-band ,
}

\DeclareAcronym{ICI}{
short= ICI,
long= inter-channel interference,
}

\DeclareAcronym{IQM}{
short= IQM,
long= in-phase and quadrature modulator,
}

\DeclareAcronym{RRC}{
short= RRC,
long= root-raised-cosine,
}

\DeclareAcronym{AWGN}{
short= AWGN,
long= additive white Gaussian noise,
}

\DeclareAcronym{SER}{
short= SER,
long= symbol error rate,
}
\DeclareAcronym{BER}{
short= BER,
long= bit error rate,
}

\DeclareAcronym{RL}{
short= RL,
long= reinforcement learning,
}

\DeclareAcronym{SNDR}{
short= SNDR,
long= signal-to-noise-plus-distortion ratio,
}

\DeclareAcronym{SNR}{
short= SNR,
long= signal-to-noise ratio,
}

\DeclareAcronym{FIR}{
short= FIR,
long= finite impulse response,
}

\DeclareAcronym{KL}{
short= KL,
long=Kullback-Leibler,
}

\DeclareAcronym{VI}{
short= VarInf,
long=variational inference,
}

\DeclareAcronym{MC}{
short= MC,
long=Monte-Carlo,
}

\DeclareAcronym{pmf}{
short= pmf,
long=conditional probability mass function,
}

\DeclareAcronym{LE}{
short= LE,
long=linear equalizer,
}

\DeclareAcronym{MMSE}{
short= MMSE,
long=minimum-mean-squared-error,
}

\DeclareAcronym{MSE}{
short= MSE,
long= mean squared error,
}
 
 \DeclareAcronym{FFE}{
short= FFE,
long=feed-forward equalizer,
}

\DeclareAcronym{ELBO}{
short= ELBO,
long= evidence lower bound,
}

\DeclareAcronym{SSMF}{
short= SSMF,
long= standard single mode fiber,
}

\DeclareAcronym{NZDSF}{
short= NZDSF,
long=non-zero dispersion-shifted fiber,
}

\DeclareAcronym{ISI}{
short= ISI,
long= inter-symbol interference,
}

\DeclareAcronym{GMP}{
short= GMP,
long= generalized memory polynomial,
}

\DeclareAcronym{MP}{
short= MP,
long= memory polynomial,
}

\DeclareAcronym{PDF}{
short= PDF,
long= probability density function,
}

\DeclareAcronym{MAP}{
short= MAP,
long=  maximum a posterior,
}

\DeclareAcronym{iid}{
short= i.i.d.,
long=  independent and identically distributed,
}

\DeclareAcronym{DBP}{
short= DBP,
long=  digital backpropagation}

\DeclareAcronym{FEC}{
short= FEC,
long=  forward error correction}

\DeclareAcronym{SPS}{
short= SPS,
long=  samples per symbol}

 \DeclareAcronym{SL}{
short= SL,
long= supervised learning}

\DeclareAcronym{DD-LMS}{
short= DD-LMS,
long=  decision-directed least-mean-squares}

\usepackage{graphics}
\usepackage{pgfplots}
\usepackage{filecontents}

\newcommand{\HW}[1]{{\color{black}{#1}}}

\title{Blind Channel Equalization Using Vector-Quantized Variational Autoencoders}
\author{%
Jinxiang Song, \emph{Student Member, IEEE},  
Vincent Lauinger, \emph{Student Member, IEEE},\\
Yibo Wu, \emph{Student Member, IEEE},
Christian H\"{a}ger, \emph{Member, IEEE},\\
Jochen Schr\"{o}der, \emph{Member, IEEE},
Alexandre Graell i Amat, \emph{Senior Member, IEEE},\\
Laurent Schmalen, \emph{Fellow, IEEE},
and Henk Wymeersch, \emph{Senior Member, IEEE}

\thanks{This work was supported by the Knut and Alice Wallenberg Foundation under grant No.~2018.0090. The work of C.~H\"{a}ger was supported by the Swedish Research Council under grant No.~2020-04718.
The work of V. Lauinger was funded by the German Federal Ministry of Education and Research under grant agreement 16KIS1316.
The work of L. Schmalen was supported by the European Research Council (ERC) under the European Union’s Horizon 2020 research and innovation program (grant agreement No. 101001899). \emph{(Corresponding author: Jinxiang Song.)}}
 
 \thanks{%
 Jinxiang Song, Christian H\"{a}ger, Alexandre Graell i Amat,  and Henk Wymeersch are with the Department of Electrical Engineering, Chalmers University of Technology, 41296 Gothenburg, Sweden (emails: \{jinxiang, christian.haeger, alexandre.graell, henkw\}@chalmers.se).}

  \thanks{Yibo Wu is with Ericsson Research, Ericsson AB, 41756 Gothenburg, Sweden, and also with the Department of Electrical Engineering, Chalmers University of Technology, 41296 Gothenburg, Sweden (e-mail: yibo.wu@ericsson.com).}
  
 \thanks{%
 Jochen Schr\"{o}der is with the Department of Microtechnology and Nanoscience, Chalmers University of Technology, 41296 Gothenburg, Sweden (email: jochen.schroeder@chalmers.se).}

\thanks{%
 Vincent Lauinger and Laurent Schmalen are with the Communications Engineering Lab (CEL), Karlsruhe Institute of Technology (KIT), 76131 Karlsruhe, Germany (email: \{vincent.lauinger, laurent.schmalen\}@kit.edu).}
}

\begin{document}
\maketitle
\begin{abstract}
State-of-the-art high-spectral-efficiency communication systems employ high-order modulation formats coupled with high symbol rates to accommodate the ever-growing demand for data rate-hungry applications. However, such systems are more vulnerable to linear and nonlinear transmission impairments, and it is important to mitigate the performance loss via digital signal processing. In this paper,  we propose a novel machine learning approach for blind channel equalization and estimation using the vector quantized (VQ) \ac{VAE} framework. The proposed approach generalizes the applicability of the conventional \ac{VAE}-based equalizer to nonlinear systems employing high-order modulation formats by introducing a codebook component and an associated novel loss function.
We evaluate the performance of the proposed method over a linear additive white Gaussian noise channel with intersymbol interference and two nonlinear scenarios. Simulation results show that the proposed method can achieve similar performance as a data aided equalizer using the \acf{MMSE} criterion, and outperforms the blind \ac{CMA} and the \ac{VAE}-based channel equalizer. Furthermore, we show that for the linear channel,  the proposed scheme exhibits better convergence properties than the \ac{MMSE}-based, the \ac{CMA}-based, and the \ac{VAE}-based equalizers in terms of both convergence speed and robustness to variations in training batch size and learning rate.
\end{abstract}


\section{Introduction}

The performance of communication systems is constrained by the linear and nonlinear distortions originating from hardware impairments and non-ideal communication channels. Such constraints become even more severe for high spectral-efficiency systems employing high-order modulation formats together with high symbol rates. To support the ever-growing demand for high-bandwidth services such as cloud computing, video streaming, and autonomous driving, it is crucial to mitigate the detrimental impact of transmission impairments that limit the systems’ throughput.

Several receiver-side \acf{DSP} techniques have been proposed for channel equalization in both wireless~\cite{malik2011adaptive} and optical communications~\cite{cartledge2017digital}. Conventionally, these algorithms either assume a channel model (e.g., a linear channel with memory~\cite{de2006accelerating}, a nonlinear channel governed by the Manakov equation\cite{luo2022learning})  or specific statistical properties in the received signal (e.g., second-order statistics\cite{tong1994blind}). In the case where such assumptions do not hold or significantly deviate from the real scenarios, these algorithms may suffer from severe performance penalties.  To circumvent these challenges,  machine learning techniques have attracted much interest in recent years, with the aim to mitigate transmission impairments directly by learning from data. In particular, \acp{NN} have been successfully applied to equalize various nonlinear channels, showing promising performance~\cite{patra1999nonlinear, kechriotis1994using,patra2009nonlinear, jarajreh2014artificial, jianping2002communication,freire2022deep, freire2022neural,hager2018nonlinear,zhang2019field}.

In general, receiver side channel equalizers fall into two categories: data aided (or pilot-based) channel equalizers and blind channel equalizers. Data aided equalizers require the transmission of  pilot symbols prior to data transmission, which decreases the overall system throughput, especially for dynamic or fast time-varying channels for which pilot symbols need to be transmitted periodically. Blind channel equalizers, on the other hand, do not require the transmission of pilot symbols and have the potential to achieve higher throughputs. Therefore, high-performance blind channel equalizers are preferable in many practical applications to achieve maximum spectral efficiency.

The most popular algorithm for blind adaptive channel equalization is the \acf{CMA}~\cite{CMA1980self} and its variants, e.g., the modified CMA~\cite{MCMA} and the multi-modulus algorithm~\cite{MMA}. The \ac{CMA} was originally designed for linear channels and phase shift keying constellations and may suffer from poor convergence when the channel is nonlinear or introduces significant distortions~\cite{nonlinear_CMA}. 
Based on the constant modulus criterion, nonlinear equalizers based on \acp{NN} have been studied in \cite{nonlinear_CMA, pandey2005feedforward,wang2009generalized}, showing improved performance compared to their linear counterparts.  Although the \ac{CMA} is immune to phase noise and has low computational complexity, it is modulation format dependent and suffers from phase ambiguity; reliable symbol detection is only possible when the \ac{CMA} is used in combination with a carrier-phase recovery block~\cite{pfau2009hardware}. The \ac{DD-LMS} algorithm~\cite{6768233,5588497} circumvents the modulation dependence of the \ac{CMA} by switching to  a decision-directed mode after proper equalizer pre-training~\cite{5464309}.

As an alternative to the \ac{CMA}-based blind channel equalizer, a different approach based on variational inference and the \acf{VAE} has been proposed in \cite{VAE_Burshtein2}. The \ac{VAE}-based blind equalizer was originally proposed for linear systems employing 4-QAM, but has later been extended to cope with a nonlinear channel considering BPSK transmission~\cite{VAE_Burshtein}. However, the nonlinear channel under consideration assumes component-wise nonlinearity only (see~\cite[Eq. (2)]{VAE_Burshtein}), and may not be able to accurately model the interplay between linear and nonlinear distortions that exist in many practical systems.
In~\cite{VAE_schmalen}, the \ac{VAE}-based equalizer was generalized to cope with high-order modulation formats coupled with probabilistic constellation shaping and shown to be more performant than the \ac{CMA}-based equalizer over a polarization-multiplexed optical fiber channel. While \cite{VAE_schmalen} showed promising performance of the \ac{VAE}-based equalizer, the approach assumes a linear dispersive channel, and the equalizer training requires a closed-form analytical solution of the \ac{ELBO} (see~\cite{VAE_Burshtein, VAE_Burshtein2, VAE_schmalen} for more details),  which is only available for channels that can be described in a simple form (e.g., the linear \ac{AWGN} channel considered in \cite{VAE_Burshtein, VAE_Burshtein2, VAE_schmalen}). However, many practical communication scenarios, including communication over the optical fiber channel, and transmission with non-ideal hardware components (e.g., nonlinear \acp{PA}, quantization-constrained receivers), cannot be accurately modeled in a simple form (e.g., as an \ac{AWGN} channel). Consequently, the applicability of the \ac{VAE}-based equalizer proposed in \cite{VAE_Burshtein, VAE_Burshtein2, VAE_schmalen} is limited.

In this paper,  we propose a novel blind channel equalizer based on a variant of the \ac{VAE}, namely the \ac{VQ-VAE}~\cite{van2017neural}. Compared to \cite{VAE_Burshtein, VAE_Burshtein2, VAE_schmalen}, training of our proposed method does not rely on the maximization of the ELBO and can be applied to both linear and nonlinear channels. It is noteworthy that in the machine learning literature, \acp{VQ-VAE} have been commonly used as generative models in fields like image, text, and audio generation. However, in the communication setup, there have been limited treatments of the \ac{VQ-VAE} framework, with only a few notable exceptions: semantic communication~\cite{hu2022robust} and joint source-channel coding~\cite{nemati2022all}. For blind channel equalization,  \acp{VQ-VAE} have never been considered before.   The main contributions of this paper are as follows: 
\begin{itemize}
	\item 1) We propose a novel blind channel equalizer based on the \ac{VQ-VAE} framework. The proposed method addresses a major shortcoming in previous works, in particular the requirement in \cite{VAE_Burshtein2, VAE_schmalen} that a closed-form expression for the \ac{ELBO} is needed for the \ac{VAE}-based equalizer training, and the fact that the nonlinear \ac{VAE} in \cite{VAE_Burshtein} is limited to BPSK transmission.

	\item 2) 	We conduct a thorough numerical study demonstrating the effectiveness of the proposed method for both a linear \ac{AWGN} channel and two nonlinear scenarios.\footnote{The complete source code to reproduce all results in this paper is available at \url{https://github.com/JSChalmers/VQ-VAE-based-equalizer}}
    We show that for the linear and nonlinear channels under investigation, the proposed method can achieve similar performance as its data-aided counterpart while outperforming the reference blind channel equalizers.
    
    \item 3) We investigate the convergence behavior of our proposed equalizer for the linear channel, showing that the proposed method exhibits better convergence properties than the chosen baseline schemes. 
\end{itemize}

The remainder of this paper is structured as follows. In Section~\ref{background}, we present our system model. In Section~\ref{preliminaries},  we start with a brief introduction to  \emph{variational inference} and the \ac{VAE}. We then briefly describe the framework of \ac{VAE}-based equalizers and its limitations. Section~\ref{VQ-VAE} describes the proposed \ac{VQ-VAE}-based blind channel equalizer and its associated cost function for training.  In Section~\ref{linear_channel} and Section~\ref{nonlinear_channel},  detailed descriptions of our simulation setup as well as numerical results are presented.  Finally, the paper is concluded in Section~\ref{conclusion}.

\emph{Notation:} $\mathbb{Z}$, $\mathbb{R}$, and $\mathbb{C}$ denote the sets of integers, real numbers, and complex numbers, respectively.
We use boldface letters to denote vectors and matrices (e.g., $\vect{x}$ and $\mat{A}$). $(\cdot)^\transpose$ and $(\cdot)^\Ctranspose$ denote transpose and conjugate transpose, respectively. 
For a vector $\smash{\vect{x} = (x_1, \ldots, x_n)^\transpose}$, $x_i$ denotes the \mbox{$i$-th} element of $\vect{x}$, $\|\vect{x}\|^2 = \sum_{i=1}^n |x_i|^2$ denotes the squared Euclidean norm.
For a matrix $\mat{A}$, $(\mat{A})_{i}$ returns the $i$--th row of $\mat{A}$, and $(\mat{A})_{i,j}$ returns the element on $i$--th row and $j$--th column of $\mat{A}$.
$\mat{I}_n$ is the $n \times n$ identity matrix.
$\mathcal{CN}(\vect{x};\boldsymbol{\mu},\vect{\Sigma})$ denotes the distribution of a proper complex Gaussian random vector with mean $\boldsymbol{\mu}$ and covariance matrix $\vect{\Sigma}$, evaluated at $\vect{x}$ ($\vect{x}$ may be omitted to represent the entire distribution). 
Lastly, $\mathbb{E}\{\cdot\}$ denotes the expected value.

\begin{figure}[t]
    \centering
     \tikzstyle{expr}=[font=\footnotesize]%
\tikzstyle{block}=[expr,rectangle, draw, thick, minimum size=30pt,
minimum height=25pt, minimum width=43pt, inner sep=3pt, rounded corners=2, fill=white]%

\tikzstyle{edge} = [draw,thick, ->, -latex]%

\begin{tikzpicture}[scale=1.2, auto, swap, every text node part/.style={align=center}]%
\def\yoff{0.7};
\def\distance{1}

	\draw[black, dashed, rounded corners=4, fill=black!5] (-5.72, 0.5) rectangle (-1.1,-0.5);%
	
	\node[] (tx0) at (-3.5, 0.8 ) {$p(\vect{y|x})$};

    \node[] (tx) at (-6.0, 0 ) {$\vect{x}$};
    
    \node[block, right= 0.35 of tx] (pulse_shaping)  {{Upsamp.}\\ {Pulse-shap.}};
    
    \node[block, right=0.3 of pulse_shaping] (channel) {{Commun.}\\ {channel} };
    
    \node[block, right=0.3 of channel] (lpf) {Filtering\\Sampling };

    \node[block, right=0.5 of lpf] (equalizer) {Equalizer};
    
    \node[block, right=0.5 of equalizer] (demapper) {Demapper};

	\node[right=0.3 of demapper] (decoded) {$\hat{\vect{x}}$};

    \draw[edge] (tx) -- (pulse_shaping);
    \draw[edge] (pulse_shaping) -- (channel);
    \draw[edge] (channel) -- (lpf);

    \draw[edge] (lpf) --node[above,pos=0.7 ]{$\vect{y}$} (equalizer);
    \draw[edge] (equalizer) --node[above,pos=0.5 ]{$\tilde{\vect{x}}$}  (demapper);
  	 \draw[edge] (demapper) -- (decoded);

\end{tikzpicture}

 
    \caption{The communication setup: discrete symbols $\vect{x}$ are upsampled and filtered by a pulse-shaping filer prior to transmission over the considered channel. At the receiver, the channel output is pre-processed (i.e., filtering, resampling) before channel equalization. The equalizer output is then used for symbol demapping. }
    \label{fig:system_model}
\end{figure}
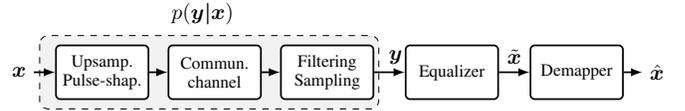

\section{System Model}

\label{background}

We consider the communication setup depicted in Fig.~\ref{fig:system_model}.  Let $\smash{\vect{x} \in \mathcal{X}^{N}}$ denote a vector of $N$ baseband symbols sampled from a complex signal constellation $\mathcal{X} \subset \mathbb{C}$, with $M=|\mathcal{X}|$ denoting the cardinality of $\mathcal{X}$. The symbols are $R$--times upsampled and filtered by a \ac{PS} filter prior to transmission over a non-ideal communication channel. In this paper, we consider block transmission, and the channel is assumed to be time-invariant within one block. Furthermore, the channels under consideration are a linear \ac{AWGN} channel with memory and two nonlinear channels, for which detailed configurations are provided in Section~\ref{linear_channel}  and Section~\ref{nonlinear_channel}, respectively. At the receiver, the channel output is filtered and resampled to have $K$ \ac{SPS}. The resulting observation vector is denoted by $\vect{y}\in \mathbb{C}^{KN}$. We remark that in this paper, the channel describes the transition from the discrete symbols $\vect{x}$ to the observation vector $\vect{y}$, and is described by the conditional probability distribution $p(\vect{y}|\vect{x})$. We further assume that the channel law $p(\vect{y}|\vect{x})$ is governed by some underlying parameters $\vect{\theta}^\star$, i.e., $p(\vect{y}|\vect{x}, \vect{\theta}^\star)$, where $\vect{\theta}^\star$ is omitted in the following for notation simplicity.

We focus on the problem of recovering the transmitted symbols $\vect{x}$ from the channel observation $\vect{y}$, for which the optimal decision is given by the \ac{MAP} criterion,
\begin{align}
    \label{Bayessian_rule}
    \hat{\vect{x}}= \underset{\vect{x}\in \mathcal{X}^N}{\arg \max} \, p(\vect{x}|\vect{y}) = \underset{\vect{x}\in \mathcal{X}^N}{\arg \max} \,\frac{p(\vect{y}|\vect{x})p(\vect{x})}{p(\vect{y})},
\end{align}
where $p(\vect{x})$ is the prior distribution of the transmitted data~$\vect{x}$, and $p(\vect{y})$ is the marginal distribution (also referred to as evidence) of the channel observations $\vect{y}$.  In practice, applying exact \ac{MAP} detection is challenging, as it requires searching over all  $M^N$ different possible input combinations, for which the computational complexity grows exponentially with the sequence length $N$ and constellation order $M=|\mathcal{X}|$.
Furthermore, the \ac{MAP} detector requires accurate knowledge of the channel (i.e., $p(\vect{y}|\vect{x})$ must be fully known). Although reliable channel estimation can be obtained at the expense of using training sequences, we consider a different setup where no channel knowledge is available at the receiver. In particular, we assume that the only information that the receiver has access to is the observation vector $\vect{y}$, the signal constellation $\mathcal{X}$, and (possibly) the structure of the underlying channel model.\footnote{For the nonlinear channel case presented in Section\ref{nonlinear_channel}, the receiver does not know the exact channel model. Instead, we assume the channel can be modeled by an \ac{NN}.}

Our goal is to design a low-complexity blind channel equalizer, so that reliable reconstruction of the transmitted symbols $\vect{x}$ can be performed from the equalizer output $\tilde{\vect{x}}$ via a (possibly) low-complexity demapper.

\section{Preliminaries}
\label{preliminaries}
In this section, we start with a brief introduction to variational inference and the \ac{VAE} framework. We then describe the \ac{VAE}-based blind channel equalizer, followed by a brief discussion of its limitations. 

\subsection{Variational Inference and Variational Autoencoder}

In variational inference, the posterior distribution $p(\vect{x}|\vect{y})$ over a set of latent variables $\vect{x}$ given some observed data $\vect{y}$ is approximated by a variational distribution $q_{\vect{\phi}}(\vect{x}|\vect{y}) \in \mathcal{Q}$, where $\mathcal{Q}$ is a family of densities over $\vect{x}$, characterized by a set of free variational parameters  $\vect{\phi}$.\footnote{Note that a realization of $\vect{y}$ is generated  according to $p(\vect{\vect{y}|\vect{x}})$ (i.e., the true channel) given some latent variables $\vect{x}$ (i.e., the transmitted symbols).} The goal of variational inference is to find the set of parameters $\vect{\phi}$ such that the variational approximation $q_{\vect{\phi}}(\vect{x}|\vect{y})$ is closest to the true posterior $p(\vect{x}|\vect{y})$ in terms of the \ac{KL} divergence~\cite{zhang2018advances},~i.e.,
\begin{align}
    \label{min_KL}
   q_{\vect{\phi}}(\vect{x}|\vect{y})= \underset{q_{\vect{\phi}}(\vect{x}|\vect{y}) \in \mathcal{Q}}{\arg \min} \mathrm{D}_{\text{KL}}(q_{\vect{\phi}}(\vect{x}|\vect{y})\Vert p(\vect{x}|\vect{y})).
\end{align}
Hence, finding the posterior distribution is cast as an optimization problem for which the complexity of the optimization is determined by the complexity of the density family $\mathcal{Q}$.

Since the true posterior $p(\vect{x}|\vect{y})$ is unknown, the \ac{KL} divergence $\mathrm{D}_{\text{KL}}(q_{\vect{\phi}}(\vect{x}|\vect{y})\Vert p(\vect{x}|\vect{y})) $ cannot be evaluated directly, thus hindering the optimization of $q_{\vect{\phi}}(\vect{x}|\vect{y})$.  However, the \ac{KL} divergence can be rewritten as
\begin{align}
\label{KL-divergence}
\begin{split}
    &\mathrm{D}_{\text{KL}}(q_{\vect{\phi}}(\vect{x}|\vect{y})\Vert p(\vect{x}|\vect{y}))\\ &= \mathrm{D}_{\text{KL}}(q_{\vect{\phi}}(\vect{x}|\vect{y})\Vert p(\vect{x}))
      - \mathbb{E}_{ q_{\vect{\phi}}(\vect{x}|\vect{y})}\left\{ \ln(p(\vect{y}|\vect{x}))\right\} + \ln (p(\vect{y})), 
\end{split}
\end{align}
where $\ln (p(\vect{y}))$ is independent of $q_{\vect{\phi}}(\vect{x}|\vect{y})$. Hence, minimizing the \ac{KL} divergence is equivalent to maximizing the \ac{ELBO} defined as 
\begin{align}
\label{ELBO}
    \text{ELBO}(\vect{\phi})=\mathbb{E}_{q_{\vect{\phi}}(\vect{x}|\vect{y})} \{\ln(p(\vect{y}| \vect{x}&))\} - \mathrm{D}_{\text{KL}}(q_{\vect{\phi}}(\vect{x}|\vect{y})\Vert p(\vect{x})),
\end{align}
for which the maximum is achieved when $q_{\vect{\phi}}(\vect{x}|\vect{y}) = p(\vect{x}|\vect{y})$.

Maximizing the \ac{ELBO} requires evaluating the expectation of the log-likelihood $\ln(p(\vect{y}|\vect{x}))$.
In many practical applications, the likelihood $p(\vect{y}|\vect{x})$ is unknown or is difficult to obtain, thus making the direct maximization of the \ac{ELBO} challenging. A promising solution for this problem has been proposed using the \ac{VAE} framework, in which an additional parametric function $p_{\vect{\theta}}(\vect{y}|\vect{x}) \in \mathcal{P}$, where $\mathcal{P}$ is a family of densities, is introduced to approximate the true likelihood $p(\vect{y}|\vect{x})$. Then, the parameter sets $\vect{\theta}$ and $\vect{\phi}$ that characterize the approximated posterior $q_{\vect{\phi}}(\vect{x}|\vect{y})$ and approximated likelihood $p_{\vect{\theta}}(\vect{y}|\vect{x})$ can be optimized simultaneously by searching over all possible combinations of $\vect{\phi}$ and $\vect{\theta}$ that maximize the approximation of the \ac{ELBO}, 
\begin{align}
	\label{eq: VAE_ELBO}
	 (\vect{\theta}, \vect{\phi})  = \underset{{\vect{\theta}\in \Theta, \vect{\phi} \in \Phi}} {\arg \max}\,  \text{ELBO}(\vect{\theta}, \vect{\phi}),
\end{align}
where 
\begin{align}
    \label{ELBO2}
    \text{ELBO}(\vect{\theta}, \vect{\phi})=\mathbb{E}_{q_{\vect{\phi}}(\vect{x}|\vect{y})} \{\ln(p_{\vect{\theta}}(\vect{y}| \vect{x}&))\} - \mathrm{D}_{\text{KL}}(q_{\vect{\phi}}(\vect{x}|\vect{y})\Vert p(\vect{x})),
\end{align}
and $\Phi$ and $\Theta$ are the search space of $\vect{\phi}$ and $\vect{\theta}$, respectively.
In the machine learning literature, $q_{\vect{\phi}}(\vect{x}|\vect{y})$ is often referred to as the inference/recognition model and $p_{\vect{\theta}}(\vect{y}|\vect{x})$ as the corresponding generative model. However, in the communication setup, we find it more intuitive to use a different terminology: we refer to $p_{\vect{\theta}}(\vect{y}|\vect{x})$ as the \emph{encoder} and to $q_{\vect{\phi}}(\vect{x}|\vect{y})$ as the corresponding \emph{decoder}.\footnote{The same terminology was used in \cite{VAE_schmalen}. In the machine learning literature, $q_{\vect{\phi}}(\vect{x}|\vect{y})$ is often referred to as the encoder while  $p_{\vect{\theta}}(\vect{y}|\vect{x})$  is the corresponding decoder.}   Typically, $q_{\vect{\phi}}(\vect{x}|\vect{y})$ and  $p_{\vect{\theta}}(\vect{y}|\vect{x})$  are implemented as \acp{NN}, where $\vect{\phi}$ and $\vect{\theta}$ are the corresponding \ac{NN} parameters~\cite{kingma2013auto}. However, it should be noted that in case a suitable model, e.g., the encoder $p_{\vect{\theta}}(\vect{y}|\vect{x})$ is available,  one can use this model instead of an \ac{NN}, as it is done in~\cite{VAE_schmalen}.

\subsection{Blind Channel Equalization Using \acp{VAE}} 
\label{VAE_equalizer}
\begin{figure}[t]
     \centering
    \includegraphics[width=1\columnwidth]{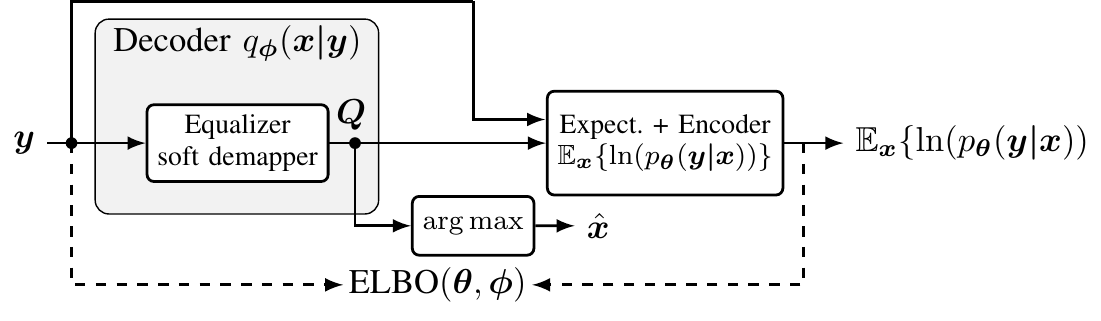}
     \caption{Blind channel equalization and estimation using \ac{VAE}, where the expectation in the expectation/encoder block is evaluated w.r.t. $\vect{x}\sim q_{\vect{\phi}}(\vect{x}|\vect{y})$. The $\arg\max$ block generates a hard decision of the transmitted message $\hat{x}_k$ according to $\smash{\hat{x}_k=\mathcal{X}_{\hat{m}_k}}$, where $\smash{\hat{m}_k =\arg\max_{m}(\vect{Q}_{k})_{m}}$, and is only used during SER evaluation. } 
     \label{fig:VAE-equalization}
 \end{figure}

Blind channel equalization (and demapping) using \acp{VAE} was originally proposed in \cite{VAE_Burshtein2}. The basic idea is to reinterpret channel equalization (and demapping) as a \emph{variational inference} problem, for which the goal is to find a parametric function (i.e., a variational approximation) that faithfully approximates the \ac{MAP} detector~\cite{VAE_Burshtein2, VAE_Burshtein, VAE_schmalen}.
To do so, as depicted in Fig.~\ref{fig:VAE-equalization}, the approach uses a \ac{VAE} decoder $\smash{q_{\vect{\phi}}(\vect{x}|\vect{y}): \mathbb{C}^{KN\times1}\to \mathbb{R}^{N\times M}}$ to generate soft estimates $\smash{\vect{Q}\in \mathbb{R}^{N\times M}}$ of the transmitted symbols $\vect{x}$ from the channel observation $\vect{y}$, and a corresponding \ac{VAE} encoder $\smash{p_{\vect{\theta}}(\vect{y}|\vect{x}): \mathbb{C}^{(K+1)N\times1}\to \mathbb{R}}$ to generate the conditional probability distribution $p_{\vect{\theta}}(\vect{y}|\vect{x})$ of receiving  $\vect{y}$ given $\smash{\vect{x}\sim q_{\vect{\phi}}(\vect{x}|\vect{y})}$ is transmitted. During the \ac{VAE} training, the expectation of the log-likelihood, $\smash{\mathbb{E}_{q_{\vect{\phi}}(\vect{x}|\vect{y})}\{\ln(p_{\vect{\theta}}(\vect{y}|\vect{x}))\}}$, is evaluated. Here, it is assumed that the transmitted symbols $\vect{x}$ are independently modulated and the \ac{ISI} caused by the channel has been mitigated. Therefore, it holds that $\smash{q_{\vect{\phi}} (\vect{x}|\vect{y}) = \prod_{k=1}^{N}q_{\vect{\phi}} (x_k |\vect{y})}$, and we define $\smash{ \vect{Q}_{k, m} \triangleq q_{\vect{\phi}} (x_k=\mathcal{X}_m |\vect{y}) }$.
The \ac{VAE} approach jointly optimizes the parameters of the \ac{VAE} encoder and decoder by maximizing \eqref{eq: VAE_ELBO}. 
Upon convergence, the \ac{VAE} decoder can be viewed as an optimized equalizer.

Since $q_{\vect{\phi}}(\vect{x}|\vect{y})$ generates soft demapping based on the channel observations, the \ac{VAE} decoder also learns a soft demapper, which can be interpreted as an approximation of the \ac{MAP} decoder. Furthermore, since the evidence $p(\vect{y})=p(\vect{y}|\vect{\theta}^\star)$ can be interpreted as the likelihood regarding the true channel parameters $\vect{\theta}^\star$, the  \ac{VAE} encoder $p_{\vect{\theta}}(\vect{y}|\vect{x})$ approximates the maximum-likelihood estimate of the channel~\cite{VAE_Burshtein, VAE_Burshtein2, VAE_schmalen}. This by-product can be used for applications such as localization~\cite{CSI_localization1, CSI_localization2} and joint communication and sensing~\cite{dorize2020capturing}.  

 \subsection{Limitations of the VAE-based Equalizer}
Training of the \ac{VAE}-based equalizer relies on maximizing the \ac{ELBO} \eqref{ELBO2}, which is commonly done by minimizing its opposite 
\begin{align}
\label{Objective}
 \mathcal{L}(\vect{\theta}, \vect{\phi}) = -\text{ELBO}(\vect{\theta}, \vect{\phi}) 
\end{align}
using  (stochastic) gradient descent, where the trainable parameters ${\vect{\lambda} = \{\vect{\theta}, \vect{\phi}\}}$ are optimized according to ${\vect{\lambda}_{t+1}=\vect{\lambda}_{t} - \eta  \nabla_{\vect{\lambda} } \mathcal{L}(\vect{\lambda}_{t})}$,
where $\eta$ is the learning rate. Assuming a linear \ac{AWGN} channel, the \ac{ELBO}, thus \eqref{Objective}, can be evaluated numerically in closed-form (see~\cite{VAE_Burshtein,VAE_Burshtein2, VAE_schmalen} or  Appendix~\ref{Append}). It is then easy to compute the gradients with respect to the trainable parameters, thereby making \ac{VAE} training straightforward (e.g, via gradient descent). 
However, for nonlinear channels that are modeled by nonlinear transfer functions, a closed-form expression for the \ac{ELBO} is generally unavailable, making the \ac{VAE} training challenging. 
In particular, to get the gradients of $\mathcal{L}(\vect{\theta}, \vect{\phi})$  with respect to the \ac{VAE} decoder parameters $\vect{\phi}$, one needs to evaluate both $\nabla_{\vect{\phi}} \mathrm{D}_{\text{KL}}(q_{\vect{\phi}}(\vect{x}|\vect{y})\Vert p(\vect{x}))$ and $\nabla_{\vect{\phi}} \mathbb{E}_{ q_{\vect{\phi}}(\vect{x}|\vect{y})}\left\{\ln(p_{\vect{\theta}}(\vect{y}|\vect{x}))\right\}$. The former is easy to compute since $\mathrm{D}_{\text{KL}}(q_{\vect{\phi}}(\vect{x}|\vect{y})\Vert p(\vect{x}))$ can often be evaluated analytically~\cite{VAE_Burshtein2}. However, evaluating $\nabla_{\vect{\phi}} \mathbb{E}_{ q_{\vect{\phi}}(\vect{x}|\vect{y})}\left\{\ln(p_{\vect{\theta}}(\vect{y}|\vect{x}))\right\}$ is more problematic due to the fact that (i) it is generally very challenging to obtain $\mathbb{E}_{ q_{\vect{\phi}}(\vect{x}|\vect{y})}\left\{\ln(p_{\vect{\theta}}(\vect{y}|\vect{x}))\right\}$ in closed-from (e.g., when $p_{\vect{\theta}}(\vect{y}|\vect{x})$ is implemented as an \ac{NN}) and (ii) the Monte-Carlo approximation of $\mathbb{E}_{ q_{\vect{\phi}}(\vect{x}|\vect{y})}\left\{\ln(p_{\vect{\theta}}(\vect{y}|\vect{x}))\right\}$ requires sampling from $q_{\vect{\phi}}(\vect{x}|\vect{y})$, for which no proper gradients are defined with respect to $\vect{\phi}$~\cite{kingma2013auto}, thus preventing the \ac{VAE} training via the gradient-based optimization methods. For continuous latent variables, an unbiased gradient approximation of $\nabla_{\vect{\phi}} \mathbb{E}_{ q_{\vect{\phi}}(\vect{x}|\vect{y})}\left\{\ln(p_{\vect{\theta}}(\vect{y}|\vect{x}))\right\}$ can be obtained using the reparameterization trick~\cite{kingma2013auto}. However, for channel equalization, where the latent variables $\vect{x}$ are discrete, the  reparameterization trick cannot be applied.

Although gradient estimators such as Reinforce~\cite{williams1992simple} or Markov-Chain-Monte-Carlo~\cite{MCMC_overview} exist in the literature, they are known to suffer from high variance or expensive computation and are therefore not considered in this paper.

\section{Blind Channel Equalization \acp{VQ-VAE} }
\label{VQ-VAE}
In this section, we first introduce the proposed framework for blind channel equalization using \acp{VQ-VAE}. We then introduce the associated loss function for the \ac{VQ-VAE}-based equalizer training.

\subsection{Blind Channel Equalization Using \acp{VQ-VAE}}

\begin{figure}[t]
    \centering
    \includegraphics[width=1\columnwidth]{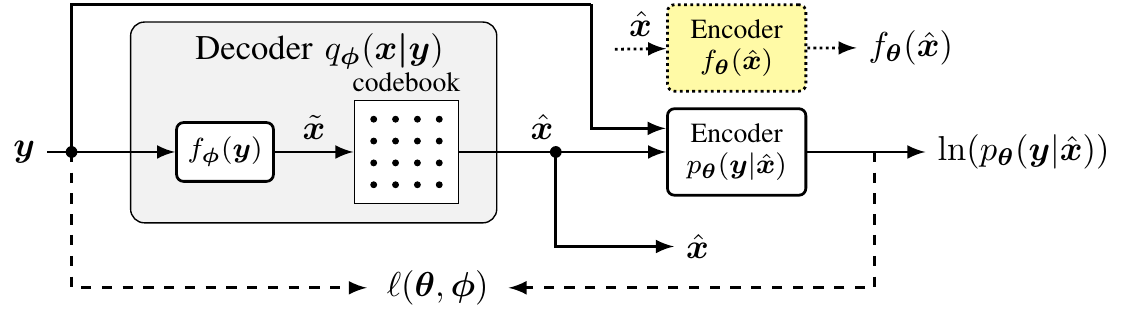}
    \caption{Blind channel equalization and estimation using \ac{VQ-VAE}. The inset (highlighted in yellow) indicates that the \ac{VQ-VAE} encoder only outputs $f_\theta(\vect{\hat{x}})$ when $p_{\vect{\theta}}(\vect{y|\hat{x}})\sim\mathcal{CN}(f_\theta(\vect{\hat{x}}), {\sigma}\vect{I}_{KN})$ is assumed.}
    \label{fig:VQ-VAE}
\end{figure}

Figure~\ref{fig:VQ-VAE} visualizes the proposed \ac{VQ-VAE}-based channel equalizer (and estimator). In addition to the encoder and the corresponding decoder of a conventional \ac{VAE}-based equalizer (see~Section~\ref{VAE_equalizer}), the proposed scheme also requires a discrete codebook component. In the machine learning literature, when the \ac{VQ-VAE} is used as a generative model, the codebook is typically unknown and is jointly trained with the encoder/decoder from data.  However, in a communication system where the latent variables (i.e., the transmitted symbols) are sampled from a known distribution, we use a fixed codebook,  which is the same as the constellation set $\mathcal{X}$. 
Similar to the conventional \ac{VAE}-based method, the \ac{VQ-VAE} decoder acts as the channel equalizer and demapper, while its corresponding \ac{VQ-VAE} encoder can be interpreted as the channel estimator.\footnote{Note that a different terminology is used in \cite{van2017neural}, where only $f_\phi(\vect{y})$ (i.e., our equalizer) is referred to as the \ac{VQ-VAE} encoder and $p_{\vect{\theta}}(\vect{y}|\hat{\vect{x}})$ is the corresponding decoder. However, we keep our terminology consistent with the \ac{VAE}-based approach~\cite{VAE_schmalen}, and we find this definition more intuitive from a communication point of view.} However, it should be highlighted that in contrast to the conventional \ac{VAE}-based method, where the \ac{VAE} decoder jointly learns channel equalization and soft demapping, our proposed \ac{VQ-VAE} decoder learns the channel equalizer $\smash{f_\phi(\vect{y}):\mathbb{C}^{KN\times1}\to \mathbb{C}^{N\times1}}$ only, while symbol demapping is performed by a following pre-defined hard demapper.  In particular, 
the channel equalizer takes the channel observation $\vect{y}$ as input and outputs a vector of equalized symbols $\tilde{\vect{x}}=f_\phi(\vect{y})$.
Then, the decoded symbol vector $\hat{\vect{x}} \in \mathcal{X}^{N}$ is obtained as
\begin{align}
    \label{quantization}
    \vect{\hat{x}} = \underset{{\vect{x} \in \mathcal{X}^{N}} }{\arg \min}\|\tilde{\vect{x}} -\vect{x}\|^2,
\end{align}
\HW{where,  similar to the conventional \ac{VAE}-based approach, it is assumed that the equalized signals are mutually independent i.e., the transmitted symbols are independently modulated and the \ac{ISI} caused by the channel memory has been mitigated by the equalizer. Note that \eqref{quantization} can be implemented with low-complexity as each equalized symbol $\tilde{{x}}_k$ can be demapped individually according to}
\begin{align}
    \label{quantization2}
    \hat{x}_k = \underset{{x \in \mathcal{X}} }{\arg \min}|\tilde{x}_k -x|^2.
\end{align}
Since the demapper considers hard demapping and is independent of the decoder parameters $\vect{\phi}$, we can express the joint transfer function $q_{\vect{\phi}}({\vect{x}}|\vect{y})$  of the  equalizer and the demapper as
\begin{align}
    \label{VQ_VAE_posterior}
    q_{\vect{\phi}}(\hat{\vect{x}}|\vect{y})= \begin{cases} 
      1 &  \hat{\vect{x}}= \underset{\vect{x} \in\mathcal{X}^{N}}{\arg \min} \|\tilde{\vect{x}} -\vect{x}\|^2 \\
      0 &  \text{otherwise},
   \end{cases}
\end{align}
which can be interpreted as a one-hot categorical distribution, and is used as the variational approximation of the \ac{MAP} decoder.
In fact, \eqref{quantization2} is essentially the widely used minimum Euclidean distance demapper, which has low complexity and is optimal for the \ac{AWGN} channel. As for the \ac{VQ-VAE} encoder $\smash{p_{\vect{\theta}}(\vect{y}|\hat{\vect{x}}): \mathbb{C}^{(K+1)N\times1}\to \mathbb{R}}$, it generates the conditional probability distribution $p_{\vect{\theta}}(\vect{y}|\hat{\vect{x}})$ of observing $\vect{y}$ given $\hat{\vect{x}}$ is transmitted.


\subsection{\ac{VQ-VAE}-based Equalizer Training}
To train the proposed \ac{VQ-VAE}-based channel equalizer and estimator, we use the following loss function
\begin{align}
    \ell(\vect{\phi}, \vect{\theta})
    &= - \mathbb{E}_{q_{\vect{\phi}} (\vect{x}|\vect{y})} \left\{\ln(p_{\vect{\theta}}(\vect{y}|\vect{x}))\right\}+ \rho \Vert \tilde{\vect{x}} - \hat{\vect{x}}\Vert ^2 \label{VQ_VAE_loss1}\\
 &= - \ln(p_{\vect{\theta}}(\vect{y}|\hat{\vect{x}}))+ \rho \Vert \tilde{\vect{x}} - \hat{\vect{x}}\Vert ^2 \label{VQ_VAE_loss2},
\end{align}
which is similar to \cite[eq.~(3)]{van2017neural} but without the term corresponding to the codebook training.
Here, \eqref{VQ_VAE_loss2} is obtained by substituting \eqref{VQ_VAE_posterior} into \eqref{VQ_VAE_loss1}, the dependence of \eqref{VQ_VAE_loss2} on $\vect{\phi}$ is implicit through $\hat{\vect{x}}$, and $\smash{\rho>0}$ is a hyperparameter that can be optimized to improve the convergence behavior of the \ac{VQ-VAE} training.  The first term on the right-hand side of \eqref{VQ_VAE_loss2} is the reconstruction loss similar to the one in \eqref{ELBO2}, but with the difference that the expectation of the log-likelihood with respect to the approximated posterior in \eqref{VQ_VAE_loss1} disappears due to \eqref{VQ_VAE_posterior}. As a result, the cumbersome requirement of computing $\nabla_{\vect{\phi}} \mathbb{E}_{ q_{\vect{\phi}}(\vect{x}|\vect{y})}\left\{\ln(p_{\vect{\theta}}(\vect{y}|\vect{x}))\right\}$ for the conventional \ac{VAE}-based equalizer training is circumvented. We highlight that even in the case where $\nabla_{\vect{\phi}} \mathbb{E}_{ q_{\vect{\phi}}(\vect{x}|\vect{y})}\left\{\ln(p_{\vect{\theta}}(\vect{y}|\vect{x}))\right\}$ for the conventional \ac{VAE}-based approach can be solved analytically, our proposed method has (i) lower computational complexity as it does not require evaluating the expectation of the log-likelihood, and (ii) better performance as we shall see in the result sections.  However, it should be noted that no gradient is defined for \eqref{quantization}. We approximate the gradient of the decoder using the straight-through estimator~\cite{bengio2013estimating} by copying the gradients from the encoder input $\hat{\vect{x}}$ to the decoder output $\vect{\tilde{x}}$.  As for the second term in \eqref{VQ_VAE_loss2}, which we refer to as the commitment loss, it drags the equalizer output to stay close to the codebook. In fact, one may find it similar to the training objective used for a \ac{DD-LMS} equalizer with a batch-wise updating scheme. However, it should be noted that the \ac{DD-LMS} equalizer requires using training overhead or the \ac{CMA} for pre-convergence, which is not the case for the proposed method.

\begin{table}[t]
\centering
\scriptsize
\caption{VAE parameters: (i) AWGN channel with linear dispersion, (ii) optical fiber channel, and (iii) AWGN channel PA nonlinearity} 
\resizebox{\columnwidth}{!}{%
\begin{tabular}{c| c ccc ccc}
\toprule
& & \multicolumn{3}{c}{VQ-VAE encoder $p_{\vect{\theta}}(\vect{y}|\vect{x})$ }   & \multicolumn{3}{c}
 { VQ-VAE decoder $q_{\vect{\phi}}(\vect{x}|\vect{y})$} \\
 \midrule
& layer    & input  & hidden& output    & input & hidden     & output  \\ 
\midrule
\multirow{3}{*}{(i)}&\#~of layers    & -  & - & -    & - & -     & -  \\ 
&\#~of neurons   & $25\times 2$ & -   & $2$      & $31\times 2$     & -      & $2$          \\ 
&act.~function   & -   & - & linear & -     & -    & linear    \\ 
\midrule
\multirow{3}{*}{(ii)}&\#~of layers    & - & $2$  & -    & - & $3$     & -  \\ 
&\#~of neurons                       & $62$     & $64/32/5$      & $2$      & $127\times 2$ & $32/8$   & $2$      \\ 
&act.~function & - & ReLU & linear & -     & ReLU    & linear    \\ 
\midrule
\multirow{3}{*}{(iii)}&\#~of layers    & - & $3$  & -    & - & $3$     & -  \\ 
&\#~of neurons    & $46$     & $64/32$      & $2$       & $95\times2$ & $64/16$   & $2$        \\ 
&act.~function & - & ReLU & linear & -     & ReLU    & linear    \\ 
\bottomrule
\end{tabular}
}
\RaggedRight

\smallskip
$\times 2$: We use real-valued \ac{NN}, and each complex number is represented by 2 neurons. For the linear channel, the \ac{VQ-VAE} is implemented as \ac{FIR} filters, and therefore there are no hidden layers.
\label{tab:network_parameters}
\end{table}

\HW{
In practice, depending on the type of data that one wants to model, it is common to assume $p_{\vect{\theta}}(\vect{y}|\vect{x})$ to be Bernoulli or Gaussian distributed~\cite{kingma2013auto}. In the communication setup, we follow the latter assumption as the noise in communication channels is commonly assumed to be Gaussian distributed. In particular, we assume $\smash{p_{\vect{\theta}}(\vect{y}|\hat{\vect{x}})\sim \mathcal{CN} (f_\theta(\hat{\vect{x}}), {\sigma}^2\vect{I}_{KN})}$, where $f_\theta(\hat{\vect{x}})$ and  ${\sigma}^2$ are generated by the \ac{VQ-VAE} encoder. It then holds that minimizing \eqref{VQ_VAE_loss2} is equivalent to minimizing~\cite{lucas2019don}
\begin{align}
    \label{training_objective}
    \ell(\theta,\phi)=\Vert \vect{y}-f_\theta(\hat{\vect{x}})\Vert ^2 + \rho \Vert \tilde{\vect{x}} - \hat{\vect{x}}\Vert ^2,
\end{align}
 since the minimum of the negative log-likelihood of a Gaussian distribution $p_{\vect{\theta}}(\vect{y}|\hat{\vect{x}})$ 
 $\sim \mathcal{CN} (f_\theta(\hat{\vect{x}}), {\sigma}^2\vect{I}_{KN})$ is always achieved at $\smash{\vect{y}=f_\theta(\hat{\vect{x}})}$.\footnote{Note that $\rho$ in \eqref{training_objective} is different from the one in \eqref{VQ_VAE_loss2}, where the former one is scaled by $\sigma^{-2}$. The effect of $\sigma^{-2}$ in \eqref{training_objective} is absorbed in $\rho$, which is a hyperparameter that we optimize for the \ac{VQ-VAE} training.} In the remainder of this paper, our \ac{VQ-VAE} encoder (shown in the inset of Fig.~\ref{fig:VQ-VAE}) generates $f_\theta(\vect{\hat{x}})$ only and is trained using \eqref{training_objective} (or its variants presented in Section~\ref{NN_VAE_realization}).}

\section{Results: Linear Channel}
\label{linear_channel}
In this section, we evaluate the proposed equalizer over an \ac{AWGN} channel with memory described by $\vect{y} = \vect{h}* \vect{x} + \vect{n}$, where $\vect{h}$ is the channel impulse response, $*$ denotes the convolution operator, and $\vect{n} \sim \mathcal{CN}(\vect{0}, \sigma_w^2\vect{I}_N)$ is \ac{iid} Gaussian noise, where $\sigma_w^2$ is the noise variance.  For the scenarios considered in this paper, all equalizers (except for the \ac{CMA}) are implemented in Python with the PyTorch framework, and  training is performed  by minimizing their corresponding cost functions using the Adam optimizer~\cite{kingma2014adam}. As for the \ac{CMA},  it is trained according to its standard implementation with one equalizer update for each processed symbol and a learning rate scheduling scheme similar to~\cite{VAE_schmalen}.

\subsection{Setup}
The considered communication setup is depicted in Fig.~\ref{fig:system_model}, where the transmitted symbols are \ac{iid} $M$-QAM symbols, the upsampling rate is $R=2$, and the \ac{RRC} filter with $10\%$ roll-off is used as the \ac{PS} filter.
 As in \cite{VAE_Burshtein, VAE_Burshtein2, VAE_schmalen}, the communication channel under consideration is a linear \ac{AWGN} channel with
\begin{align*}
    \vect{h} =  (0.055 + j0.05, 0.283 -j0.120, -0.768 + j0.279,\\ -0.064-j0.058, 0.047-j0.023),
\end{align*}
where the noise variance of the \ac{AWGN} is chosen from a range of considered \acp{SNR}, which we define as $\mathbb{E}\{|x_k|^2\}/\sigma_w^2$. Since $\smash{R=2}$ is used as the upsampling rate and we consider operating the proposed equalizer on 2 \ac{SPS}, no resampling is performed prior to channel equalization.

 \begin{figure}[t]
    \centering
    \includegraphics[width=1\columnwidth]{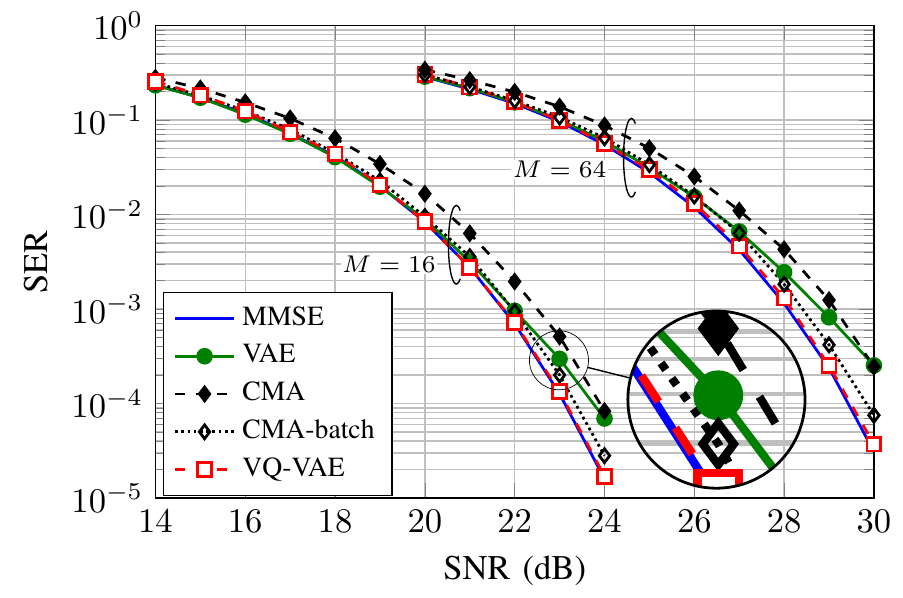}
    \caption{SER performance over a linear \ac{AWGN} channel with finite memory, a separate equalizer is trained for each data point shown in the figure. }
    \label{fig:AWGN_SER}
\end{figure}

 \begin{figure*}[t]
    \centering
    \vspace*{-3mm}
    \includegraphics[width=1\textwidth]{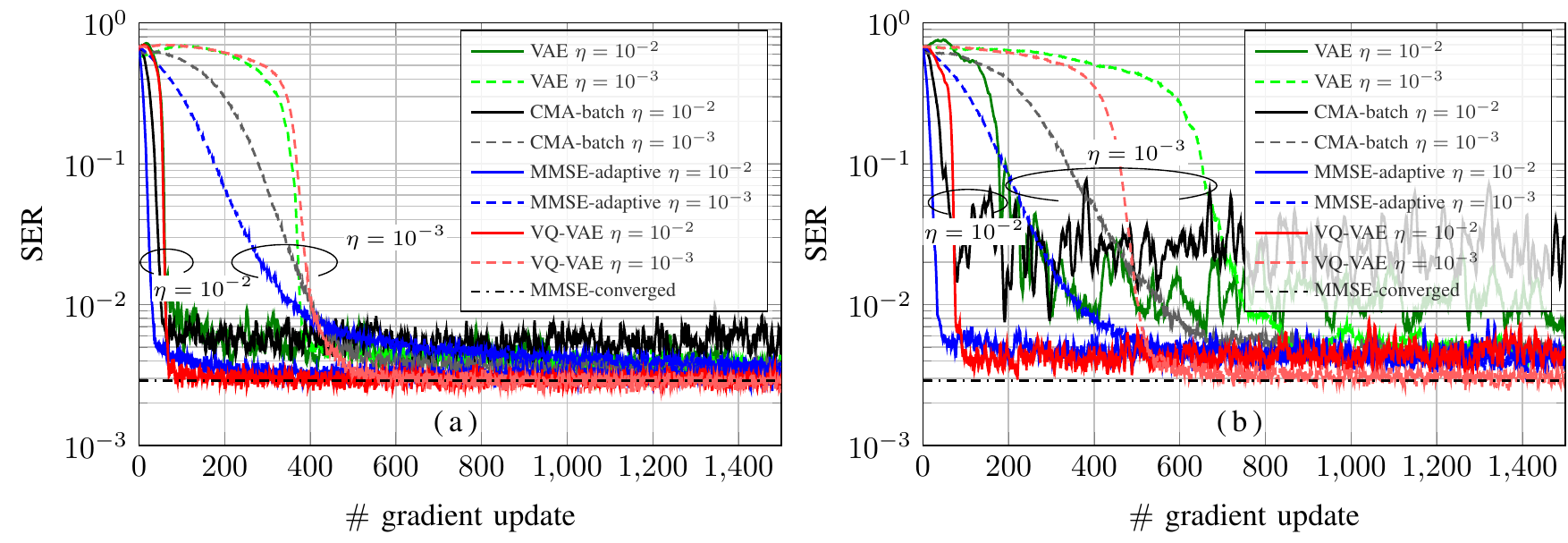}
    \vspace*{-8mm}
    \caption{ Convergence behavior of four channel equalizers for 16-QAM transmission at $\text{SNR}= 21\,\text{dB}$: (a) each gradient update utilizes $N=1024$ symbols; (b) each gradient update utilizes $N=64$ symbols.}
    \label{fig:convergence_test}
\end{figure*}

\subsubsection{VQ-VAE realization and training}
Similar to \cite{VAE_schmalen}, we implement both the \ac{VQ-VAE} encoder and decoder as linear \ac{FIR} filters with Dirac initialization. However, we highlight again thatin contrast  to~\cite{VAE_schmalen}, our proposed equalizer does not require a soft demapping block at the training stage and therefore has lower computational complexity.\footnote{In case a soft demapping block is required to cope with the soft-decision \ac{FEC}, a standard Gaussian demapper can be used once the equalizer has converged.}
We note that one can also implement the channel equalizer as an \ac{NN} (see~\cite{VAE_Burshtein,VAE_Burshtein2, VAE_schmalen}). However, the \ac{NN}-based equalizer in general requires more trainable parameters but does not bring performance improvement for the linear channel case~\cite[Fig.~7]{VAE_schmalen}, and is therefore not considered here. The detailed \ac{VQ-VAE} configurations are summarized in Table~\ref{tab:network_parameters}, and we train our proposed \ac{VQ-VAE}  by minimizing \eqref{VQ_VAE_loss3} with $\smash{\rho=1}$ using stochastic gradient descent; in Section~\ref{impact_of_hyperparameter}, we study the impact of $\rho$ on the \ac{VQ-VAE} training. The training dataset contains $2^{16}$ symbols, the batch size is $\smash{N=1024}$ (corresponding to 2048 samples), and the learning rate is $\smash{\eta=0.001}$. The reference equalizers (except the CMA) are trained with the same batch size (i.e., $\smash{N=1024}$). However, for each reference equalizer,  the learning rate is optimized within the set $\{1, 2, 5, 0.1, 0.2, 0.5, 0.01, 0.02, 0.05\}\times 10^{-2}$, i.e., we use the $\eta$ that achieves the best performance.

 \subsection{Results and Discussions} 
 
 \subsubsection{SER versus SNR}
 The \ac{SER} achieved by the proposed \ac{VQ-VAE} is shown in Fig.~\ref{fig:AWGN_SER}, where a separate equalizer is trained for each data point. For comparison, we also show the \ac{SER} achieved by the conventional \ac{VAE}-based  and the \ac{CMA}-based channel equalizer. The \ac{CMA} is phase insensitive and an additional phase recovery block is required prior to symbol detection. In this paper, we consider a genie-aided phase recovery, so that no performance penalty is introduced in the phase-recovery block. Moreover, we provide the performance that can be achieved by a data aided \ac{FFE}, for which the mean squared error is chosen as the training criterion. Upon convergence, the data aided \ac{FFE} resembles the \ac{MMSE} equalizer, and we therefore do not distinguish these two equalizers in the following. For both 16-QAM and 64-QAM, the proposed method achieves the same performance as the data aided \ac{FFE}, while outperforming the conventional \ac{VAE}-based and the \ac{CMA}-based equalizer, especially at high \acp{SNR}. One possible explanation for the performance gap between the conventional \ac{VAE} and the \ac{VQ-VAE} is that the soft demapping block in the conventional \ac{VAE} requires the variance of the \ac{AWGN} as input, which may lead to a mismatched decoding criterion in case where  residual \ac{ISI} remains or the estimated noise variance does not represent the real one sufficiently well.\footnote{\HW{One possible solution for mitigating the performance loss due to the mismatched soft demapping can be found in \cite{schmalen2017performance}, where the authors propose to obtain the noise variance from the mutual information estimate.}}

 In order to close the performance gap between the \ac{CMA} and the proposed equalizer, we implement a blind equalizer using a batch-wise updating scheme, for which the constant modulus criterion is used as the cost function for training.\footnote{Although the standard \ac{CMA} implementation updates the filter taps for each processed symbol, it is also common to use batch-wise updating scheme to enable parallelized data processing (see~e.g., \cite{parallel_CMA, parallel_CMA2}). }  The \ac{SER} of the resulting equalizer, which we denote by CMA-batch, is also shown in Fig.~\ref{fig:AWGN_SER}.  It can be seen that the CMA-batch outperforms the standard \ac{CMA} and the conventional \ac{VAE} (at high \acp{SNR}), but does not perform as well as the data aided \ac{FFE} or the proposed \ac{VQ-VAE}-based approach. We attribute the remaining performance gap between CMA-batch and the data aided \ac{FFE} to the fact that the constant modulus criterion is ill-conditioned for high-order modulation formats.

\subsubsection{Convergence behavior}
We now study the convergence behavior of our proposed method and compare it with the reference schemes. For all equalizers under investigation, we consider 16-QAM transmission at $\text{SNR}=21\,\text{dB}$ with training data generated on the fly (i.e., new training data is generated after each gradient update). Figure~\ref{fig:convergence_test} shows the \ac{SER} evolution of different equalizers as we increase the number of performed gradient updates. The number of symbols used for each gradient update is $\smash{N=1024}$ for Fig.~\ref{fig:convergence_test}\,(a) and $\smash{N=64}$ for Fig.~\ref{fig:convergence_test}\,(b). For the considered configurations,  we observe that: (i) A large batch size in general leads to better convergence behavior for all studied equalizers, which is expected as a larger batch size results in a smaller variance in the gradient estimate. However, it should be noted that a large batch size is more computationally demanding and requires slower variations in the channel. There exists an ``optimal'' value that trades off the performance against the computational complexity and convergence speed. (ii) The proposed \ac{VQ-VAE}-based approach is more robust to variations in batch size and learning rate. In particular, for different batch sizes and learning rates under investigation, the proposed method always achieves the best \ac{SER} performance among all considered equalizers upon convergence, which we believe to be advantageous for time-varying channels. (iii) For a large batch size $N=1024$, a large learning rate generally leads to faster convergence for all considered equalizers, except that the \ac{CMA}-batch suffers from training instability. The proposed equalizer has very similar convergence behavior as the conventional \ac{VAE}, while having a slightly slower starting convergence rate than the CMA-batch and the data aided \ac{FFE}. However, it should be highlighted that among all considered equalizers, the proposed approach requires fewest gradient updates to converge to the ``optimal'' performance. (iv) For a small batch size $\smash{N=64}$, the proposed equalizer is able to converge to the ``optimal'' performance, while all reference equalizers exhibit a certain level of \ac{SER} penalty, especially for a relatively large learning rate $\eta=10^{-2}$. 


 \begin{figure}[t]
    \centering
    \includegraphics[width=1\columnwidth]{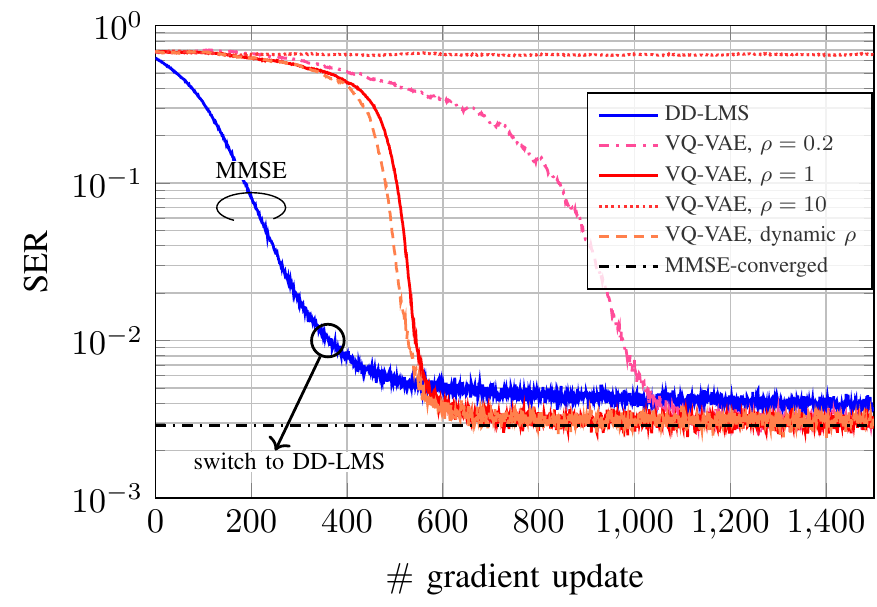}
    \caption{Convergence behavior of the \ac{VQ-VAE}-based and \ac{DD-LMS} equalizer for 16-QAM transmission at $\text{SNR}=21\,\text{dB}$. The learning rate is $\eta=10^{-2}$ and the number of symbols used for each gradient update is $N=64$ for both the \ac{VQ-VAE}-based and \ac{DD-LMS} equalizer. }
    \label{fig:convergence_study2}
\end{figure}


\subsubsection{Impact of $\rho$ and comparison with DD-LMS}
\label{impact_of_hyperparameter}
We now study the impact of the hyperparameter $\rho$ on the \ac{VQ-VAE} training. Figure~\ref{fig:convergence_study2} visualizes the \ac{SER} as a function of the number of gradient updates. It is shown that for $\smash{\rho=0.2}$, the \ac{VQ-VAE}-based channel equalizer requires more gradient updates to convergence as compared to $\smash{\rho=1}$. However, for $\smash{\rho=10}$, the \ac{VQ-VAE}-based equalizer fails to converge. Such a behavior is expected due to the fact that when $\rho$ is set to a large value, the proposed equalizer resembles a \ac{DD-LMS} equalizer with batch-wise updating scheme, and will inevitably suffer from convergence issues if no proper pre-training is performed.
To avoid the need for optimizing $\rho$ manually,  we modify \eqref{training_objective} to the loss function
\begin{align}
    \label{VQ_VAE_loss3}
    \ell(\vect{\phi}, \vect{\theta}) =\psi_{t}\Vert \vect{y}- f_
    \theta(\hat{\vect{x}})\Vert ^2 + (1-\psi_{t}) \Vert \tilde{\vect{x}} - \hat{\vect{x}}\Vert ^2,
\end{align}
 where $t > 0$ is the iteration index and
\begin{align}
    \label{dynamic_weighting}
	\psi_{t+1} = \frac{\Vert \vect{y}- f_
    \theta(\hat{\vect{x}})\Vert ^2}{\Vert \vect{y}- f_
    \theta(\hat{\vect{x}})\Vert ^2 + \Vert \tilde{\vect{x}} - \hat{\vect{x}}\Vert ^2}\bigg|_{t},
\end{align}
with $\psi_{0}=0.5$ and $f(\cdot)|_{t}$ indicates evaluating $f(\cdot)$ at $t$--th iteration. This new loss function dynamically balances the importance of the reconstruction error and the commitment loss. The learning curve obtained using \eqref{dynamic_weighting} is shown in Fig.~\ref{fig:convergence_study2}.  Last but not least, we show the learning curve of a \ac{DD-LMS} equalizer, for which the equalizer taps  are pre-conditioned in a data aided manner using the \ac{MMSE} criterion. Once the \ac{SER} reaches $10^{-2}$, we switch the equalizer to the decision-directed mode. The resulting \ac{SER} evolution is shown in Fig.~\ref{fig:convergence_study2}, where it is shown that the \ac{DD-LMS} equalizer does not perform as well as the proposed \ac{VQ-VAE}-based equalizer.

\section{Results: Nonlinear Channels}
\label{nonlinear_channel}

To validate the effectiveness of the proposed method for nonlinear channels,  we evaluate our proposed equalizer over two nonlinear scenarios namely (i) transmission over an unamplified optical fiber channel and (ii) transmission over a wireless channel with nonlinear \ac{PA}. Since both fiber and \ac{PA} nonlinearity are highly dependent on the transmit power, for all results presented in the following, a separate equalizer is trained for each power level.

\subsection{Nonlinear Optical Fiber Channel}
\label{optical-channel}
We start with evaluating the performance of the proposed method over an unamplified optical fiber link, for which typical use cases include data center interconnects with a range of $80-120\,\text{km}$.  To support transmission over $80\,\text{km}$ and beyond, such systems are required to operate at a (relatively) high launch power and often exhibit moderate/high fiber nonlinearity.

\begin{table}[t]
\setlength{\tabcolsep}{0.6em}
\centering
\caption{Fiber parameters}
\resizebox{1\columnwidth}{!}{%
\begin{tabular}{ccc ccc }
\toprule
\multicolumn{3}{c}{SSMF} &\multicolumn{3}{ c}{NZDSF} \\
 \midrule
  $\alpha$ (dB/km)   & $\beta$ ($\mathrm{ps^2/km}$) & $\gamma$ $\mathrm{(W\cdot km)}^{-1}$  &  $\alpha$ (dB/km)   & $\beta$ ($\mathrm{ps^2/km}$) & $\gamma$ $\mathrm{(W\cdot km)}^{-1}$ \\ 
\midrule
$0.2$  & $-21.683$ & 1.3 & $0.21$  & $-4.0$ & $1.6$  \\ 
\bottomrule
\end{tabular}
}
\label{tab:fiber_parameters}
\end{table}

 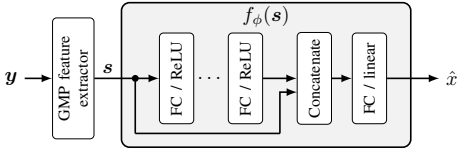
\begin{figure}[t]
    \centering
    \tikzstyle{expr}=[font=\footnotesize]%
\tikzstyle{block}=[expr,rectangle, draw, minimum size=45pt,
minimum height=17pt, inner sep=3pt, rounded corners=2, fill=white]%
\tikzstyle{edge} = [draw, thick, ->, -latex]%
\begin{tikzpicture}[scale=1.2, auto, swap, every text node part/.style={align=center}]%

        \draw[black, solid, thick, rounded corners=4, fill=black!5] (-0.8, 1.1) rectangle (3.4, -1);%
      \node[] () at (1.3, 0.9) {$f_\phi(\vect{s})$};

        \node[rotate=0] (input) at (-2.4,0) {$\vect{y}$};   
        \node[block, rotate=90,  minimum width=60pt] (mapper) at (-1.5,0) {\footnotesize{GMP feature}\\\footnotesize{extractor}};
        \draw[edge] (input) -- (mapper); 


        \node[rotate=0] (input) at (-1.0, 0.15) {$\vect{s}$};    
        \node[block, rotate=90] (layer1) at (0,0) {\footnotesize{FC / ReLU}};
        \node[] () at (0.5,0) {$\cdots$};

         \draw[edge] (mapper) -- (layer1);

        \node[block, rotate=90] (layer2) at (1,0) {\footnotesize{FC / ReLU}};
        \node[block, rotate=90] (cat) at (2,0) {\footnotesize{Concatenate}};
    
        \draw[edge] (layer2) -- (cat);  
        \node[block, rotate=90] (output) at (2.8,0) {\footnotesize{FC / linear}};
        \draw[edge] (cat) -- (output); 
        \node[] (equalized) at (4,0) {$\hat{x}$};
        \draw[edge] (output) -- (equalized); 
        
        \draw[edge, -] (-0.6, 0) -- (-0.6, -0.8); 
        \draw[edge, -] (-0.6, -0.8) -- (1.5, -0.8); 
	\draw[edge, rounded corners=0pt] (1.5, -0.8) |- (1.75, -0.2);
	\node at (-0.6,0) [circle,fill,inner sep=1.2pt]{};
 
\end{tikzpicture}%
    \caption{Structure of the \ac{NN}-based channel equalizer employing fully connected architecture and a residual connection that connects the \ac{NN} input directly to the output layer. The feature extractor block generates the nonlinear feature vector as described in Section~\ref{NN_VAE_realization}.}
    \label{fig:NN_structure}
\end{figure}

\subsubsection{Setup} The considered system transmits a single channel waveform of $25\,\text{GBaud}$ over a single span fiber link of $110\,\text{km}$.  The \ac{PS} filter is the \ac{RRC} filter with $10\%$ roll-off and the oversampling rate is set to $R=8$ to ensure that the spectral broadening effect due to fiber nonlinearity is fully captured in the simulation.  To cover different scenarios, we consider two test cases with different level of fiber nonlinearity: (i) transmission over the \ac{SSMF} and (ii) transmission over the \ac{NZDSF}.  The optical signal propagation along the fiber is simulated by solving the nonlinear Schr\"{o}dinger equation via the split-step Fourier method (SSFM) with 100 steps.\footnote{No improvement of accuracy for the forward propagation was observed when further increasing the number of steps or the oversampling rate.} The fiber parameters used in the simulation are summarized in Table~\ref{tab:fiber_parameters}. At the receiver, \ac{AWGN} is added to the fiber output to emulate the shot noise from the coherent receiver. Similar to \cite{goossens2022introducing}, we consider a coherent receiver with a minimum sensitivity of $-20\,\text{dBm}$, and the amount of noise that is added to the fiber output signal is $-33.5\,\text{dBm}$. In a back-to-back setup, such a configuration will allow for achieving a pre-\ac{FEC} \ac{BER} of $1.25\cdot 10^{-2}$, which is the maximum \ac{BER} that can be tolerated by the  \ac{FEC} block in the 400ZR standard~\cite{400NR} to fulfill the post decoding error rate requirements. The resulting waveform is then low-pass filtered by a brick-wall filter with $45\,\text{GHz}$ bandwidth and sampled at a sampling rate of $50\,\text{GHz}$ (i.e., to have $2$ \ac{SPS}), after which a static filter is applied to perform a coarse (i.e., $90\%$ in our simulations) dispersion compensation.\footnote{Here, we note that the dispersion compensation filter can in principle be integrated into the adaptive channel equalizer. However, the required equalizer length in such an implementation will increase linearly with the fiber length. In practice, it is more common to use a static filter to perform a coarse dispersion compensation prior to the adaptive filtering.} Finally, channel equalization is performed prior to symbol detection.

 \begin{figure*}[t]
    \centering
    \includegraphics[width=1\textwidth]{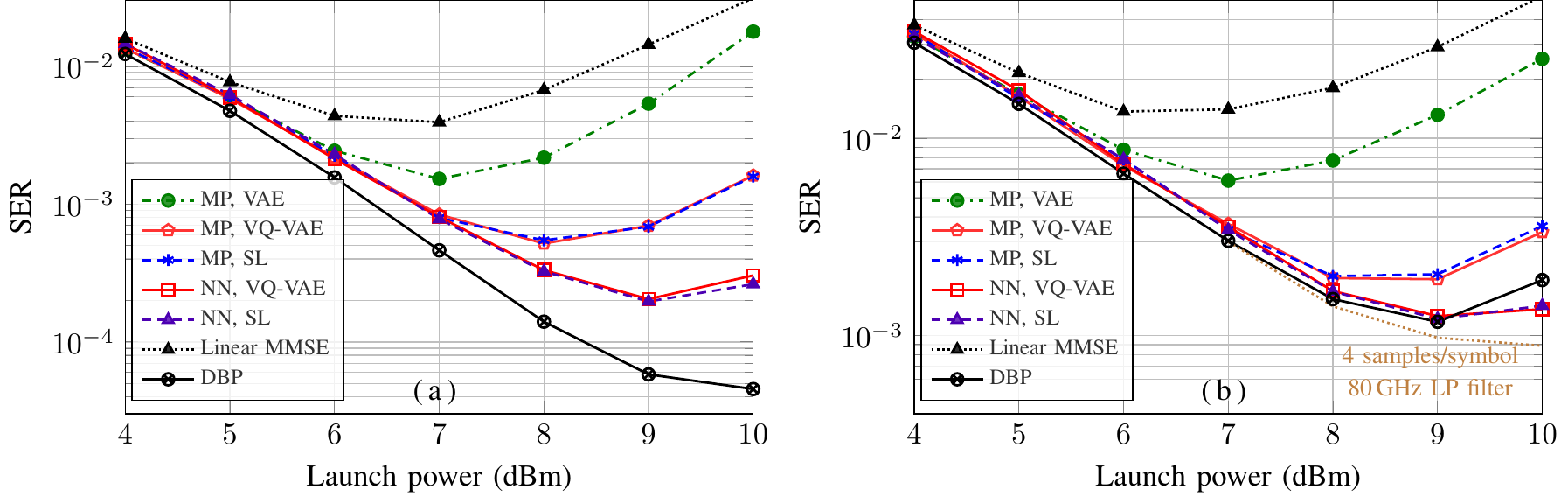}
    \vspace*{-7mm}
    \caption{SER as a function of launch power for transmission over an unamplified optical fiber link of 110 km: (a) standard single mode fiber with $90\%$ dispersion pre-compensated; (b) non-zero dispersion-shifted fiber with $90\%$ dispersion pre-compensated. Note that the DBP curve assumes performing the SSFM for 100 steps, which has very high complexity.}
    \label{fig:fiber_channel_ser}
\end{figure*}

\subsubsection{VQ-VAE realization and training} 
\label{NN_VAE_realization}
Traditional methods using linear \ac{FIR} filters cannot model or equalize the nonlinear optical fiber channel well. \acp{NN} on the other hand the ability to learn nonlinear and complex relationships in data. Therefore, we implement our VQ-VAE using a pair of \acp{NN}. In the literature, various \ac{NN} architectures ranging from simple multi-layer perceptrons~\cite{patra1999nonlinear} to sophisticated models such as biLSTM~\cite{freire2022deep} have been used for channel equalization. In this paper, we restrict ourselves to the fully-connected architecture due to its easy deployment and training. However, for the decoder \ac{NN} (i.e., our channel equalizer),  a residual connection that directly connects the input feature to the output layer is also employed as depicted in Fig.~\ref{fig:NN_structure}. Such a residual connection has been shown to improve the performance as well as \ac{NN} training in several communication related applications~\cite{bajaj2022deep,wu2021low,8555649}. Furthermore, instead of having the raw channel observations $\vect{y}$ as input,  our \ac{NN}-based equalizer takes a nonlinear feature vector $\vect{s}^k$ as input and generates the $k$--th equalized signal according to  $\smash{\tilde{x}_k = f_\phi(\vect{s}^{k})}$. Here, 
the nonlinear feature vector $\vect{s}_k$ is constructed following the \ac{GMP}~\cite{morgan2006generalized}, which is a widely used model that relates the desired output (i.e., equalized signal $\tilde{x}_k$) to the input (i.e., observation $\vect{y}$) according to
\begin{IEEEeqnarray}{rCl}
     \tilde{x}_k &=& \sum_{p\in \mathcal{P}_a}\sum_{l\in \mathcal{L}^p_a} a_{p,l}\,\underbrace{y_{k-l}|y_{k-l}|^{p-1}}_{s^{k}_{p,l}}\nonumber\\
       &&+\!\!\sum_{p\in \mathcal{P}_b}\sum_{l\in \mathcal{L}^p_b} \sum_{m\in \mathcal{M}_b} b_{p,l,m}\,\underbrace{y_{k-l}|y_{k-l-m}|^{p-1}}_{s^{k}_{p,l,m}}\nonumber\\
       &&+\!\!\sum_{p\in \mathcal{P}_c}\sum_{l\in \mathcal{L}^p_c} \sum_{n\in \mathcal{M}_c}\,\underbrace{c_{p,l,m}\,y_{k-l}|y_{k-l+m}|^{p-1}}_{s^k_{p,l,n}},\IEEEeqnarraynumspace
\end{IEEEeqnarray}
where $a_{p,l}$, $b_{p,l,m}$, and $c_{p,l,m}$ are the model coefficients,  $\mathcal{P}_a$, $\mathcal{P}_b$, and $\mathcal{P}_c$ are the sets that contain the nonlinear orders, $\mathcal{L}^p_a$, $\mathcal{L}^p_b$, $\mathcal{L}^p_c$, $\mathcal{M}_b$, and $\mathcal{M}_c$ are the sets that contain the memory taps. 
In our simulation, the nonlinear feature vector $\vect{s}^{k}$ contains all $s^k_{p,l}$, $s^{k}_{p,l,m}$, and $s^k_{p,l,n}$ for ${\mathcal{P}_a =\{1,3,5\}}$, ${\mathcal{P}_b =\mathcal{P}_c=\{3\}}$, ${\mathcal{L}^{1}_a =\{-21,\ldots,21\}}$, ${\mathcal{L}^{3}_a =\{-5,\ldots,5\}}$, ${\mathcal{L}^{5}_a =\{-3,\ldots,3\}}$, $\mathcal{L}_b=\mathcal{L}_c =$ $\{5,\ldots,5\}$, and ${\mathcal{M}_b=\mathcal{M}_c =\{1,2,3\}}$. With this configuration, the constructed nonlinear features vector is of length $127$. Our \ac{NN}-based equalizer, summarized in Table~\ref{tab:network_parameters}, is found to perform better than the one using a fully-connected architecture.\footnote{A similar approach has been applied in \cite{kamalov2018evolution, zhang2019field}, where nonlinear impairment features constructed from the first-order perturbation theory were used as the input to the NN-based equalizer. Also note that our configuration is only optimized for one launch power (i.e., $8\,\text{dBm}$). To further improve the \ac{SER} performance or to reduce complexity, one may need to optimize the \ac{NN} configuration (e.g., number of layers, \ac{NN} architeture) for each launch power.}

The training of our proposed \ac{VQ-VAE} is again performed by minimizing \eqref{training_objective}, but with slight modifications. 
In particular, we use $L_2$ regularization
on the trainable weights in the input layer and the hidden layers of our equalizer \ac{NN}. A similar regularization scheme has been previously used in~\cite{freire2022neural}, where it is shown that using a regularization term in the loss function can help to mitigate the so-called ``jail window'' effect observed in the equalized signal constellation when the regression loss is used for training~\cite{diedolo2022nonlinear}. Note that this jail window constellation can lead to a performance penalty unless it is handled properly~\cite{freire2022neural,diedolo2022nonlinear}.
For optimization,  our training dataset contains $65536$ symbols, and the batch size and learning rate are set to $\smash{N=2048}$ and $\smash{\eta=0.001}$, respectively.


 \begin{figure}[t]
    \centering
    \includegraphics[width=1\columnwidth]{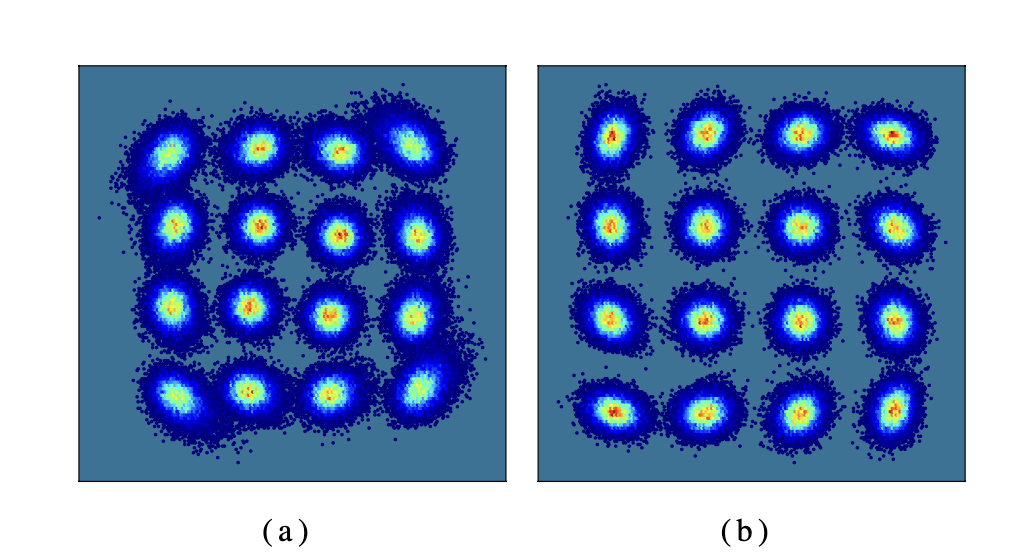}

    \caption{Received signal constellations after channel equalization (a): Linear equalizer; (b) NN-based equalized trained with VQ-VAE. The symbol rate of the transmitted 16-QAM signal is $25\,\text{GBaud}$ and the signal is propagated over the \ac{SSMF} of 110km with a launch power of $8\,\text{dBm}$.}
    \label{fig:equalized_const}
\end{figure}

 \subsubsection{SER versus launch power} 
The \ac{SER} (shown in red squares) achieved by the proposed \ac{VQ-VAE}-based equalizer over a range of launch power is shown in Fig.~\ref{fig:fiber_channel_ser}\,(a) for the \ac{SSMF} and in Fig.~\ref{fig:fiber_channel_ser}\,(b) for the \ac{NZDSF}, respectively. For comparison, we also provide the \ac{SER} achieved by an \ac{NN}-based equalizer with the same \ac{NN} configurations but trained using \ac{SL} (purple triangles),  a data aided linear equalizer, and \ac{DBP} with 100 steps. It can be seen that for a low launch power (e.g., at 4 dBm), the performance gap between the linear equalizer and the nonlinear one is small, indicating that the \ac{SER} is mostly dominated by the receiver noise. However, when increasing the launch power, the nonlinear distortion gets stronger, and the performance gap between the linear and nonlinear equalizer increases.  In particular, the optimal transmit power for the linear equalizer is around 7\,dBm, and is increased to around 9\,dBm for the NN-based nonlinear equalizer. The increased optimal launch power indicates that systems employing the proposed nonlinear equalizer can support a longer transmission distance, which is essential for the unamplified optical fiber link.

In order to show the advantages of the proposed \ac{VQ-VAE}-based equalizer over the \ac{VAE}-based approach for the nonlinear channels, we derived a closed-form expression (see~Appendix~\ref{Append2}) for the \ac{ELBO} assuming that the nonlinear channel can be modeled by a \ac{MP}.\footnote{ Although a closed-form expression for the \ac{ELBO} is available in this case, we note that the computational complexity of the \ac{VAE}-based approach is much higher than the \ac{VQ-VAE}-based method.} When implementing both the encoder and decoder as \acp{MP}, the \ac{SER} achieved by the \ac{MP}-based \ac{VAE} (green circles) and the \ac{MP}-based \ac{VQ-VAE} (red pentagons) is also shown in Fig.~\ref{fig:fiber_channel_ser}, where it can be seen that the proposed \ac{VQ-VAE}-based equalizer significantly outperforms the \ac{VAE}-based counterpart  at high launch power.
 
Finally, it is noteworthy that the \ac{NN}-based equalizer slightly outperforms \ac{DBP} in the nonlinear regime for the \ac{NZDSF}, although the optimal SER is essentially the same. A similar behavior was observed in~\cite{hager2018nonlinear}, where it was justified that in the nonlinear regime, a significant part of the broadened spectrum of the received waveform is filtered out before processing. The missing spectrum is not properly backpropagated when using \ac{DBP}, while the \ac{NN}-based equalizer manages to somewhat compensate for that effect. Indeed, as we increase the bandwidth of the receiver and perform \ac{DBP} with 4 \ac{SPS}, the performance of \ac{DBP} improves and is better than that of the \ac{NN}-based equalizer.
  
Figure~\ref{fig:equalized_const} visualizes the equalized signal obtained by using the linear \ac{FFE} (left) and the proposed VQ-VAE-based equalizer with NN implementation (right) assuming transmission over the \ac{SSMF} with a launch power of $8$\,dBm. Clearly, the signal constellation obtained using the NN-based equalizer is much less degraded compared to the one obtained using the linear \ac{FFE}, albeit nonlinear distortion can still be observed even when the \ac{NN}-based equalizer is applied. To further improve the performance of the \ac{NN}-based equalizer, 
one may consider using different \ac{NN} architectures such as the biLSTM~\cite{freire2022deep} or the physics-informed one~\cite{hager2018nonlinear}. Here, we highlight that the proposed \ac{VQ-VAE}-based equalizer is topology independent, and we do not focus on designing the optimal architecture of \ac{NN}-based channel equalizers. Yet, we highlight that a hyperparameter optimization is performed in our simulation to find a low-complexity topology that achieves satisfying performance.

\subsection{Nonlinear Wireless Channel with PA Nonlinearity}
\label{PA-channel}

We now study a wireless scenario, where both linear and nonlinear distortion are introduced to the transmit signal by the non-ideal \ac{PA}. We note that in the literature, it is customary to compensate for \ac{PA} impairments at the transmitter side using \ac{DPD}\cite{kim2001digital}. However, to optimize the \ac{DPD} parameters, a feedback data acquisition path is typically required to collect the \ac{PA} output signal\cite{liu2018beam}. To capture the full-band behavior of the \ac{PA}, which is necessary for the \ac{DPD} optimization, high sampling rate analog-to-digital converters are needed for the feedback path, which is challenging and expensive for wideband signals. Additionally, for scenarios where low-cost and low-power consumption devices are preferred, e.g., the internet of things, transmitter-side \ac{DPD} optimization may not be affordable. Therefore, we study the possibility of mitigating the transmitter-induced performance loss at the receiver-end using blind channel equalization.

\subsubsection{Setup}
The \ac{PA} is modeled by a \ac{GMP}, for which the nonlinear orders, the memory lengths, and the cross-term memory lengths are $\smash{\mathcal{P}_a=\mathcal{P}_b=\mathcal{P}_c=\{1,2,3,4,5,6,7\}}$, $\mathcal{L}^{p}_a=\mathcal{L}^{p}_b=\mathcal{L}^{p}_c=$ $\{0,1,2\}$ for all $p$, and $\mathcal{M}_b=\mathcal{M}_c=$  $\{1\}$, respectively. The corresponding parameters of the \ac{GMP} are estimated from the measurements of the RF WebLab,\footnote{The RF WebLab is a PA measurement setup that can be remotely accessed at \url{www.dpdcompetition.com}} for which more detailed configurations can be found in \cite{PA_model}. Similar to~\cite{PA_model}, we consider the transmission of a 64-QAM signal over an \ac{AWGN} channel, where the upsampling rate is $R=2$, the \ac{PS} filter is an \ac{RRC} filter with $20\%$ roll-off, and the \ac{AWGN} has a fixed noise standard deviation $\sigma_\text{ch}=0.4\,\text{V}$. We vary the \ac{PA} output power to emulate different operating SNRs.

\subsubsection{Equalizer realization and training} Similar to the nonlinear fiber channel case,  the proposed \ac{VQ-VAE} is implemented as a pair of \acp{NN}, which have a very similar structure (but different hyper-parameters) as the ones used in Section~\ref{optical-channel}. In particular, the equalizer \ac{NN} also employs a residual connection (see~Fig.~\ref{fig:NN_structure}) and the \ac{NN} input $\vect{s}^{k}$ is constructed as described in Section~\ref{NN_VAE_realization}, where the model parameters are set to ${\mathcal{P}_a =\{1,\ldots,7\}}$, ${\mathcal{P}_b =\mathcal{P}_c=\{3\}}$, ${\mathcal{L}^{1}_a =\{-15,\ldots,15\}}$, ${\mathcal{L}^{p}_a =\{-3,\ldots,3\}}$ for $p\neq1$,  ${\mathcal{L}_b=\mathcal{L}_c =\{-3,\ldots,3\}}$, and $\mathcal{M}_b=\mathcal{M}_c =\{1,2,3\}$.
The \ac{NN} parameters are summarized in Table~\ref{tab:network_parameters}.  
As for training, we noticed that the nonlinear \ac{PA} scenario requires using a larger batch size to achieve satisfying performance compared to the case of an unamplified optical fiber link.  We attribute it to the fact that we now work with 64-QAM transmission and we operate at a lower \ac{SER}.  In our simulation, the training dataset contains $2^{18}$ symbols, the batch size is set to $2^{14}$, and the learning rate is $\eta=0.001$.

\begin{figure}[t]
    \centering
    \includegraphics[width=1\columnwidth]{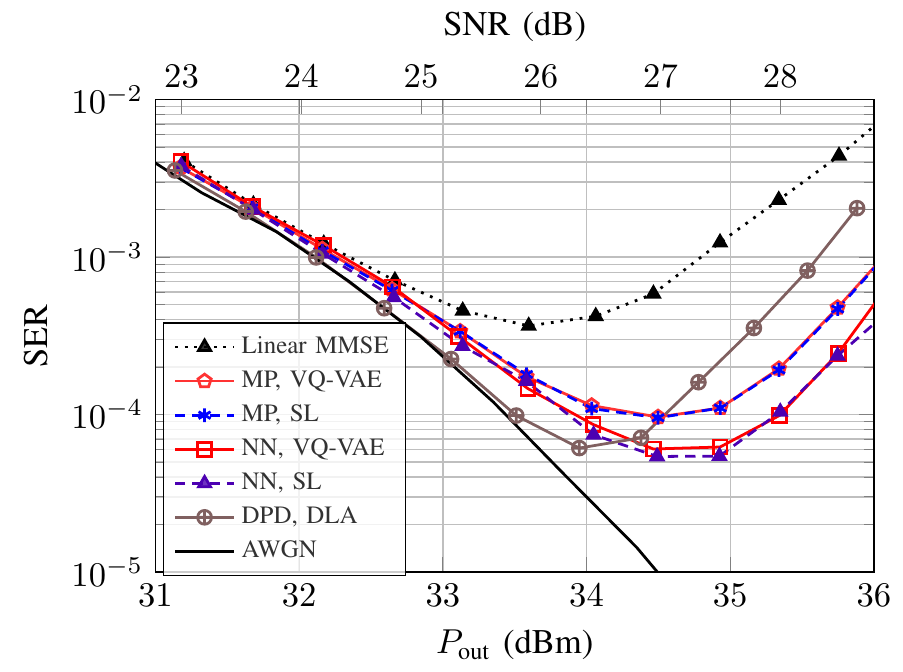}
    \caption{SER performance an AWGN channel where the transmitted signals are distorted by a nonlinear PA.}
    \label{fig:PA_SER}
\end{figure}

\subsubsection{SER versus averaged PA output power} 
Figure~\ref{fig:PA_SER} shows
the achieved SER as a function of the average \ac{PA} output power for the case of using a linear equalizer, the \ac{VQ-VAE}-based equalizer (NN-based and MP-based), and the theoretical SER of 64-QAM over an \ac{AWGN}. We observe that when the average \ac{PA} output power is low, most of the input signal to the  \ac{PA} lies in the \ac{PA}'s linear region, and increasing the \ac{PA} output power leads to improved \ac{SER} performance (as the \ac{SNR} is increased).  As the average \ac{PA} output power exceeds 33\,dBm, the \ac{PA} starts to operate in the medium-high nonlinear region, and the SER starts to increase when the receiver uses a linear equalizer. In the case where the proposed nonlinear equalizer is used, a substantial performance gain can be achieved compared to using a linear equalizer. In particular, the optimal average \ac{PA} output power can be increased by around 1.5\,dBm (from 33.5\,dBm to 35\,dBm), while the optimal SER is improved from $4\cdot10^{-4}$ to $5.5\cdot10^{-5}$. 

To compare the proposed method with the customary implementation where the \ac{PA} nonlinearity is compensated at the transmitter side, the \ac{SER} achieved by a \ac{DL}-based \ac{DPD}, implemented as an \ac{NN}, is also provided~\cite{directlearning_DPD}. It is shown that the proposed equalizer achieves essentially the same performance as a \ac{DL}-based \ac{DPD} in terms of \ac{SER}. However, the \ac{DL}-based \ac{DPD} is shown to outperform the \ac{VQ-VAE}-based equalizer at low lunch power. We attribute the performance gap partially to the fact that the \ac{VQ-VAE}-based equalizer is trained with data distorted by the \ac{AWGN}, which is not the case for the \ac{DPD}.  As for high launch power (e.g., above $34\,\text{dBm}$), the proposed method significantly outperforms the \ac{DL}-based \ac{DPD}. We believe the performance gap comes from the fact that the input signal (after DPD) to the \ac{PA} exhibits higher peaks and is therefore more distorted by the nonlinear \ac{PA}.


\section{Conclusion}
\label{conclusion}
In this work, we proposed a novel approach to blind channel equalization using the \ac{VQ-VAE} framework. We evaluate the performance of the proposed method over both a linear  channel and two nonlinear scenarios including transmission over the unamplified optical fiber link and transmission over a wireless channel with \ac{PA} nonlinearity.  For the linear channel, the proposed method, where both the \ac{VQ-VAE} encoder and decoder are implemented as linear \ac{FIR} filters,  is shown to achieve similar performance to a data aided equalizer using the \ac{MMSE} criterion, while outperforming state-of-the-art blind equalizers including the \ac{CMA} and \ac{VAE}-based equalizers. Compared to the reference schemes, the proposed equalizer is also shown to exhibit better convergence behavior, which can be advantageous for time-varying channels. Moreover, contrary to the \ac{VAE}-based approach that is  designed for linear \ac{AWGN} channels, we have shown that our proposed method can also converge for nonlinear channels, which is advantageous for many practical communication systems where nonlinear impairments are presented.



\appendices
\section{Closed-form Expression for the ELBO for a linear \ac{AWGN} channel}

\label{Append}
Consider the linear \ac{AWGN} channel described by $\vect{y} = \vect{h}* \vect{x} + \vect{n}$, where $\vect{h}$ is the channel impulse response, $*$ denotes the convolution operator, and $\vect{n} \sim \mathcal{CN}(\vect{0}, \sigma_w^2\vect{I}_N)$ is \ac{iid} Gaussian noise with variance $\sigma_w^2$. The channel law (likelihood) $p_{\vect{\theta}}(\vect{y}|\vect{x})$ follows the multivariate Gaussian distribution 
\begin{align}
    \label{likelihood_linear}
    p_{\vect{\theta}}(\vect{y}|\vect{x}) = \frac{1}{(\pi \sigma_w^2)^N}\, e^{-\Vert\vect{y} - \vect{X}\vect{h}\Vert^2/{\sigma_w^2}},
\end{align}
where $\theta=\{\vect{h}, \sigma_w^2\}$, $\vect{X}\vect{h} = \vect{h}*\vect{x}$, and
\begin{align}
    \vect{X} = (\vect{x}^{(L)},\ldots, \vect{x}^{(1)}, \vect{x},\vect{x}^{(-1)},\ldots, \vect{x}^{(-L)}),
\end{align}
where $\smash{\vect{x}^{(l)} = (x_{1+l},\ldots, x_{N}, \vect{0}_{l}^\top)^\top}$, $\vect{0}_{l}$ denotes a length $l$ column vector of all zero elements, and $\vect{x}^{(-l)} = (\vect{0}_{l}, x_{1},\ldots, x_{N-l})^\top$. 
It is assumed that the channel has $(2L+1)$ taps and $y_k$ is related to $\{x_{k-L},\ldots,x_{k+L}\}$.
Assuming the transmitted symbols are uniformly distributed (i.e., $p(\vect{x})=M^{-N}$) and the outputs of the \ac{VAE} equalizer are mutually independent (i.e., $q_{\vect{\phi}}(x_i, x_j|\vect{y}) =q_{\vect{\phi}}(x_i|\vect{y})q_{\vect{\phi}}(x_j|\vect{y})$ for $i\neq j$), we have
\begin{IEEEeqnarray}{rCl}\label{linear_ELBO}
    \text{ELBO}(\vect{\phi}, \vect{\theta})
    &=&\mathbb{E}_{q_{\vect{\phi}}(\vect{x}|\vect{y})} \{\ln(p(\vect{y}| \vect{x}))\} - \mathrm{D}_{\text{KL}}(q_{\vect{\phi}}(\vect{x}|\vect{y})\Vert p(\vect{x})) \nonumber\\
    &=& -\>N \ln (\pi\sigma_w^2) - \frac{1}{\sigma_w^2} \mathbb{E}_{q_{\vect{\phi}}(\vect{x}|\vect{y})}\{\Vert\vect{y}-\vect{Xh}\Vert^2\}\nonumber\\  &&+ \mathcal{H}(q_{\vect{\phi}}(\vect{x}|\vect{y}))+ N\ln(M)\nonumber\\
    &=& -\>N \ln (\sigma_w^2) -\frac{1}{\sigma_w^2} A + \mathcal{H}(q_{\vect{\phi}}(\vect{x}|\vect{y})) +C,
\end{IEEEeqnarray}
where $\mathcal{H}(q_{\vect{\phi}}(\vect{x}|\vect{y}))$ is the entropy of $q_{\vect{\phi}}(\vect{x}|\vect{y})$, $C$ is a constant independent of $\vect{\phi}$ and $\vect{\theta}$, and
\begin{align*}
\begin{split}
    \mathrm{A} &=\mathbb{E}_{q_{\vect{\phi}}(\vect{x}|\vect{y})}\{\Vert\vect{y}-\vect{Xh} \Vert^2\}\\
    &= \vect{y}^{\dagger}\vect{y} -  2\Re\{\vect{y}^\dagger \mathbb{E}_{q_{\vect{\phi}}(\vect{x}|\vect{y})}\{\vect{Xh}\}\} + \mathbb{E}_{q_{\vect{\phi}}(\vect{x}|\vect{y})}\{(\vect{Xh})^\dagger \vect{Xh} \}\\
  & =  \vect{y}^{\dagger}\vect{y} -  2\Re\{\vect{y}^\dagger \mathbb{E}_{q_{\vect{\phi}}(\vect{x}|\vect{y})}\{\vect{X}\}\vect{h}\} + \vect{h}^\dagger \mathbb{E}_{q_{\vect{\phi}}(\vect{x}|\vect{y})}\{\vect{X}^\dagger \vect{X} \}\vect{h},
\end{split}
\end{align*}
where
\begin{align*}
\begin{split}
     &\mathbb{E}_{q_{\vect{\phi}}(\vect{x}|\vect{y})}\left\{\vect{X}^\dagger \vect{X}\right\}_{i,j}  \\&= \begin{cases} 
      \mathbb{E}_{q_{\vect{\phi}}(\vect{x}|\vect{y})}\left\{\Vert\vect{x}^{(L-i)}\Vert^2\right\} & i=j \\
      \mathbb{E}_{q_{\vect{\phi}}(\vect{x}|\vect{y})}\left\{(\vect{x}^{(L-i)})^\dagger  \right\}
      \mathbb{E}_{q_{\vect{\phi}}(\vect{x}|\vect{y})}\left\{\vect{x}^{(L-j)}\right\} &  i\neq j. 
   \end{cases}
\end{split}
\end{align*}

\section{Closed-form expression for the ELBO for a nonlinear channel modeled by MP}

\label{Append2}

 Consider modeling the nonlinear channel using an \ac{MP}\cite{morgan2006generalized} according to
 \begin{align}
     y_k =\sum_{p \in \mathcal{P}}  \sum_{l\in \mathcal{L}^{p}}c_{p,l}\,x_{k-l}|x_{k-l}|^p + n_k,
 \end{align}
 where $\mathcal{P}$ and $\mathcal{L}^p$ are the sets containing the nonlinear orders and memory lengths, $x_k$ and $y_k$ are the input and output to the channel at the $k$--th time step,  $n_k \sim \mathcal{CN}(0, \sigma_w^2)$ is the \ac{AWGN},  and $c_{m,k}$ denotes model coefficient, respectively. We can relate the channel output to the input in a compact matrix representation given by
 \begin{align*}
     \vect{y} = \vect{X}\vect{c} + \vect{n},
 \end{align*}
 where $\vect{n}\sim \mathcal{CN}(\vect{0}, \sigma_w^2\vect{I}_N)$ and
 \begin{align*}
     \mat{X}^\top =\begin{bmatrix}
      (\vect{x}^{(L)})^\top\\
     \vdots\\
     (\vect{x}^{(-L)})^\top\\
     \vdots\\ (\vect{x}^{(L)}|\vect{x}^{(L)}|^P)^\top\\
     \vdots\\ (\vect{x}^{(-L)}|\vect{x}^{(-L)}|^P)^\top
     \end{bmatrix},\, 
     \text{and}\,\, \vect{c} = \begin{bmatrix}
      c_{0, L}\\
     \vdots\\
     c_{0,-L}\\
     \vdots\\ c_{P,L}\\
     \vdots\\ c_{P,-L}
     \end{bmatrix},
 \end{align*}
 where, for notation convenience, it is assumed $\mathcal{L}^{p}=\{-L,\ldots,L\}$ for all $p\in \mathcal{P}=\{1,\ldots, P\}$.
 Then, the likelihood has the same form as \eqref{likelihood_linear}, except that the convolution $\vect{h}*\vect{x}$ should be replaced by $\mat{X}\vect{c}$. Similarly, the ELBO can be evaluated in closed-form and is given by \eqref{linear_ELBO}, but with
 \begin{align*}
 \begin{split}
     \mathrm{A} &=\mathbb{E}_{q_{\vect{\phi}}(\vect{x}|\vect{y})}\{\Vert\vect{y-}\vect{Xh}\Vert^2 \}\\
   & = \vect{y}^{\dagger}\vect{y} -  2\Re\{\vect{y}^\dagger \mathbb{E}_{q_{\vect{\phi}}(\vect{x}|\vect{y})}\{\vect{X}\}\vect{h}\} + \vect{h}^\dagger \mathbb{E}_{q_{\vect{\phi}}(\vect{x}|\vect{y})}\{\vect{X}^\dagger \vect{X} \}\vect{h},
 \end{split}
 \end{align*}
 and
 \begin{align*}
 \begin{split}
      &\mathbb{E}_{q_{\vect{\phi}}(\vect{x}|\vect{y})}\{\vect{X}^\dagger \vect{X}\}_{i,j}=\\ &\begin{cases} 
       \mathbb{E}_{q_{\vect{\phi}}(\vect{x}|\vect{y})}\{(\vect{x}^{(L-a)}|\vect{x}^{(L-a)}|^e)^\dagger (\vect{x}^{(L-b)}|\vect{x}^{(L-b)}|^f) \} & a= b \\
       \mathbb{E}_{q_{\vect{\phi}}(\vect{x}|\vect{y})}\{((\vect{x}^{(L-a)}|\vect{x}^{(L-a)}|^e)^\dagger \} \\
       \indent \indent \indent \indent \times \mathbb{E}_{q_{\vect{\phi}}(\vect{x}|\vect{y})}\{(\vect{x}^{(L-a)}|\vect{x}^{(L-a)}|^f)\} & a\neq b,
   \end{cases}
 \end{split}
 \end{align*}
 where $a, b\equiv i,j \bmod (2L+1)$,  $e=\lfloor \frac{i}{2L+1} \rfloor$, and $f=\lfloor \frac{j}{2L+1} \rfloor$.

\section*{Acknowledgment}
Jinxiang Song greatly appreciates the Ericsson Research Foundation for supporting his research visit to the Communications Engineering Lab (CEL) at Karlsruhe Institute of Technology (KIT), 76131 Karlsruhe, Germany.

\balance
\bibliographystyle{IEEEtran}
\bibliography{references} 

\begin{thebibliography}{10}
\providecommand{\url}[1]{#1}
\csname url@samestyle\endcsname
\providecommand{\newblock}{\relax}
\providecommand{\bibinfo}[2]{#2}
\providecommand{\BIBentrySTDinterwordspacing}{\spaceskip=0pt\relax}
\providecommand{\BIBentryALTinterwordstretchfactor}{4}
\providecommand{\BIBentryALTinterwordspacing}{\spaceskip=\fontdimen2\font plus
\BIBentryALTinterwordstretchfactor\fontdimen3\font minus
  \fontdimen4\font\relax}
\providecommand{\BIBforeignlanguage}[2]{{%
\expandafter\ifx\csname l@#1\endcsname\relax
\typeout{** WARNING: IEEEtran.bst: No hyphenation pattern has been}%
\typeout{** loaded for the language `#1'. Using the pattern for}%
\typeout{** the default language instead.}%
\else
\language=\csname l@#1\endcsname
\fi
#2}}
\providecommand{\BIBdecl}{\relax}
\BIBdecl

\bibitem{malik2011adaptive}
G.~Malik and A.~S. Sappal, ``Adaptive equalization algorithms: An overview,''
  \emph{Int. J. Adv. Comput. Sci. Appl.}, vol.~2, no.~3, 2011.

\bibitem{cartledge2017digital}
J.~C. Cartledge, F.~P. Guiomar, F.~R. Kschischang, G.~Liga, and M.~P. Yankov,
  ``Digital signal processing for fiber nonlinearities,'' \emph{Opt. Express},
  vol.~25, no.~3, pp. 1916--1936, 2017.

\bibitem{de2006accelerating}
S.~Qureshi, ``Adaptive equalization,'' \emph{Proc. IEEE}, vol.~73, no.~9, pp.
  1349--1387, 1985.

\bibitem{luo2022learning}
S.~Luo, S.~K.~O. Soman, L.~Lampe, J.~Mitra, and C.~Li, ``Learning for
  perturbation-based fiber nonlinearity compensation,'' in \emph{Proc. Eur.
  Conf. Opt. Commun (ECOC)}, 2022, p. We1C.4.

\bibitem{tong1994blind}
L.~Tong, G.~Xu, and T.~Kailath, ``Blind identification and equalization based
  on second-order statistics: A time domain approach,'' \emph{IEEE Trans. Inf.
  Theory}, vol.~40, no.~2, pp. 340--349, 1994.

\bibitem{patra1999nonlinear}
J.~C. Patra, R.~N. Pal, R.~Baliarsingh, and G.~Panda, ``Nonlinear channel
  equalization for {QAM} signal constellation using artificial neural
  networks,'' \emph{IEEE Trans. Syst., Man, Cybern., B}, vol.~29, no.~2, pp.
  262--271, 1999.

\bibitem{kechriotis1994using}
G.~Kechriotis, E.~Zervas, and E.~S. Manolakos, ``Using recurrent neural
  networks for adaptive communication channel equalization,'' \emph{IEEE Trans.
  Neural Netw.}, vol.~5, no.~2, pp. 267--278, 1994.

\bibitem{patra2009nonlinear}
J.~C. Patra, W.~C. Chin, P.~K. Meher, and G.~Chakraborty,
  ``Legendre-{FLANN}-based nonlinear channel equalization in wireless
  communication system,'' in \emph{Proc. IEEE Int. Conf. Syst., Man, Cybern.},
  2008, pp. 1826--1831.

\bibitem{jarajreh2014artificial}
M.~A. Jarajreh \emph{et~al.}, ``Artificial neural network nonlinear equalizer
  for coherent optical {OFDM},'' \emph{IEEE Photon. Technol. Lett.}, vol.~27,
  no.~4, pp. 387--390, 2014.

\bibitem{jianping2002communication}
D.~Jianping, N.~Sundararajan, and P.~Saratchandran, ``Communication channel
  equalization using complex-valued minimal radial basis function neural
  networks,'' \emph{IEEE Trans. Neural Netw.}, vol.~13, no.~3, pp. 687--696,
  2002.

\bibitem{freire2022deep}
P.~J. Freire, J.~E. Prilepsky, Y.~Osadchuk, S.~K. Turitsyn, and V.~Aref, ``Deep
  neural network-aided soft-demapping in coherent optical systems: Regression
  versus classification,'' \emph{IEEE Trans. Commun.}, vol.~70, no.~12, pp.
  7973--7988, 2022.

\bibitem{freire2022neural}
P.~J. Freire, A.~Napoli, B.~Spinnler, N.~Costa, S.~K. Turitsyn, and J.~E.
  Prilepsky, ``Neural networks-based equalizers for coherent optical
  transmission: Caveats and pitfalls,'' \emph{IEEE J. Sel. Top. Quantum
  Electron.}, vol.~28, no.~4, pp. 1--23, 2022.

\bibitem{hager2018nonlinear}
C.~H{\"a}ger and H.~D. Pfister, ``Nonlinear interference mitigation via deep
  neural networks,'' in \emph{Proc. Opt. Fiber Commun. Conf. (OFC)}, 2018, p.
  W3A.4.

\bibitem{zhang2019field}
S.~Zhang \emph{et~al.}, ``Field and lab experimental demonstration of nonlinear
  impairment compensation using neural networks,'' \emph{Nat. Commun.},
  vol.~10, no.~1, pp. 1--8, 2019.

\bibitem{CMA1980self}
D.~Godard, ``Self-recovering equalization and carrier tracking in
  two-dimensional data communication systems,'' \emph{IEEE Trans. Commun.},
  vol.~28, no.~11, pp. 1867--1875, 1980.

\bibitem{MCMA}
K.~N. Oh and Y.~O. Chin, ``Modified constant modulus algorithm: blind
  equalization and carrier phase recovery algorithm,'' in \emph{Proc. IEEE Int.
  Conf. Commun. (ICC)}, vol.~1, 1995, pp. 498--502.

\bibitem{MMA}
J.~Yang, J.-J. Werner, and G.~A. Dumont, ``The multimodulus blind equalization
  and its generalized algorithms,'' \emph{IEEE J. Sel. Areas Commun.}, vol.~20,
  no.~5, pp. 997--1015, 2002.

\bibitem{nonlinear_CMA}
D.~Wang and D.~Wang, ``Comparison of nonlinear {CMA}-criterion-based blind
  equalizers,'' in \emph{Proc. IEEE Int. Conf. on Wireless Commun. Netw. and
  Mobile Comput. (WiCom)}, 2010.

\bibitem{pandey2005feedforward}
R.~Pandey, ``Feedforward neural network for blind equalization with {PSK}
  signals,'' \emph{Neural. Comput. Appl.}, vol.~14, no.~4, pp. 290--298, 2005.

\bibitem{wang2009generalized}
D.~Wang and D.~Wang, ``Generalized derivation of neural network constant
  modulus algorithm for blind equalization,'' in \emph{Proc. IEEE Int. Conf.
  Wireless Commun., Netw. Mobile Comput.}, 2009.

\bibitem{pfau2009hardware}
T.~Pfau, S.~Hoffmann, and R.~No{\'e}, ``Hardware-efficient coherent digital
  receiver concept with feedforward carrier recovery for {M}-{QAM}
  constellations,'' \emph{J. Lightw. Technol.}, vol.~27, no.~8, pp. 989--999,
  2009.

\bibitem{6768233}
R.~W. Lucky, ``Techniques for adaptive equalization of digital communication
  systems,'' \emph{Bell Syst. Tech. J.}, vol.~45, no.~2, pp. 255--286, 1966.

\bibitem{5588497}
M.~S. Faruk, Y.~Mori, C.~Zhang, and K.~Kikuchi, ``Proper polarization
  demultiplexing in coherent optical receiver using constant modulus algorithm
  with training mode,'' in \emph{Proc. Optoelectron. Commun. Conf. (OECC)},
  2010, pp. 768--769.

\bibitem{5464309}
S.~J. Savory, ``Digital coherent optical receivers: Algorithms and
  subsystems,'' \emph{IEEE J. Sel. Top. Quantum Electron.}, vol.~16, no.~5, pp.
  1164--1179, 2010.

\bibitem{VAE_Burshtein2}
A.~Caciularu and D.~Burshtein, ``Blind channel equalization using variational
  autoencoders,'' in \emph{Proc. IEEE Int. Conf. Commun. Workshops (ICC
  Workshops)}, 2018.

\bibitem{VAE_Burshtein}
------, ``Unsupervised linear and nonlinear channel equalization and decoding
  using variational autoencoders,'' \emph{IEEE Trans. Cogn. Commun. Netw.},
  vol.~6, no.~3, pp. 1003--1018, 2020.

\bibitem{VAE_schmalen}
V.~Lauinger, F.~Buchali, and L.~Schmalen, ``Blind equalization and channel
  estimation in coherent optical communications using variational
  autoencoders,'' \emph{IEEE J. Sel. Areas Commun.}, vol.~40, no.~9, pp.
  2529--2539, 2022.

\bibitem{van2017neural}
A.~Van Den~Oord, O.~Vinyals \emph{et~al.}, ``Neural discrete representation
  learning,'' \emph{Adv. Neural Inf. Process. Syst. (NIPS)}, vol.~30, 2017.

\bibitem{hu2022robust}
Q.~Hu, G.~Zhang, Z.~Qin, Y.~Cai, G.~Yu, and G.~Y. Li, ``Robust semantic
  communications with masked {VQ-VAE} enabled codebook,'' \emph{arXiv preprint
  arXiv:2206.04011}, 2022.

\bibitem{nemati2022all}
M.~Nemati and J.~Choi, ``All-in-one: {VQ-VAE} for end-to-end joint
  source-channel coding,'' 2022.

\bibitem{zhang2018advances}
C.~Zhang, J.~B{\"u}tepage, H.~Kjellstr{\"o}m, and S.~Mandt, ``Advances in
  variational inference,'' \emph{IEEE Trans. Pattern Anal. Mach. Intell.},
  vol.~41, no.~8, pp. 2008--2026, 2018.

\bibitem{kingma2013auto}
D.~P. Kingma and M.~Welling, ``Auto-encoding variational bayes,'' \emph{arXiv
  preprint arXiv:1312.6114}, 2013.

\bibitem{CSI_localization1}
S.~Abdul~Samadh, Q.~Liu, X.~Liu, N.~Ghourchian, and M.~Allegue, ``Indoor
  localization based on channel state information,'' in \emph{Proc. IEEE Top.
  Conf. Wireless Sensors and Sensor Netw. (WiSNeT)}, 2019.

\bibitem{CSI_localization2}
M.~A. Al-Qaness \emph{et~al.}, ``Channel state information from pure
  communication to sense and track human motion: A survey,'' \emph{Sensors},
  vol.~19, no.~15, p. 3329, 2019.

\bibitem{dorize2020capturing}
C.~Dorize, S.~Guerrier, E.~Awwad, and J.~Renaudier, ``Capturing acoustic speech
  signals with coherent {MIMO} phase-{OTDR},'' in \emph{Proc. Eur. Conf. Opt.
  Commun. (ECOC)}, 2020.

\bibitem{williams1992simple}
R.~J. Williams, ``Simple statistical gradient-following algorithms for
  connectionist reinforcement learning,'' \emph{Machine learning}, vol.~8,
  no.~3, pp. 229--256, 1992.

\bibitem{MCMC_overview}
C.~Karras, A.~Karras, M.~Avlonitis, and S.~Sioutas, ``An overview of {MCMC}
  methods: from theory to applications,'' in \emph{Proc. Int. Conf. on
  Artificial Intell. Appl. Innov.}\hskip 1em plus 0.5em minus 0.4em\relax
  Springer, 2022, pp. 319--332.

\bibitem{bengio2013estimating}
Y.~Bengio, N.~L{\'e}onard, and A.~Courville, ``Estimating or propagating
  gradients through stochastic neurons for conditional computation,''
  \emph{arXiv preprint arXiv:1308.3432}, 2013.

\bibitem{lucas2019don}
J.~Lucas, G.~Tucker, R.~B. Grosse, and M.~Norouzi, ``Don't blame the {ELBO}!
  {A} linear {VAE} perspective on posterior collapse,'' \emph{Adv. Neural Inf.
  Process. Syst. (NeurIPS)}, vol.~32, 2019.

\bibitem{kingma2014adam}
D.~P. Kingma and J.~Ba, ``Adam: A method for stochastic optimization,''
  \emph{arXiv preprint arXiv:1412.6980}, 2014.

\bibitem{schmalen2017performance}
L.~Schmalen, A.~Alvarado, and R.~Rios-M{\"u}ller, ``Performance prediction of
  nonbinary forward error correction in optical transmission experiments,''
  \emph{J. Lightw. Technol.}, vol.~35, no.~4, pp. 1015--1027, 2017.

\bibitem{parallel_CMA}
D.~E. Crivelli \emph{et~al.}, ``Architecture of a single-chip 50 {G}b/s
  {DP}-{QPSK}/{BPSK} transceiver with electronic dispersion compensation for
  coherent optical channels,'' \emph{IEEE Trans. Circuits Syst. I, Reg.
  Papers}, vol.~61, no.~4, pp. 1012--1025, 2014.

\bibitem{parallel_CMA2}
N.~Kaneda and A.~Leven, ``Coherent polarization-division-multiplexed {QPSK}
  receiver with fractionally spaced {CMA} for {PMD} compensation,'' \emph{IEEE
  Photon. Technol. Lett.}, vol.~21, no.~4, pp. 203--205, 2009.

\bibitem{goossens2022introducing}
S.~Goossens \emph{et~al.}, ``Introducing 4d geometric shell shaping for
  mitigating nonlinear interference noise,'' \emph{J. Lightw. Technol.},
  vol.~41, no.~2, pp. 599--609, 2023.

\bibitem{400NR}
\BIBentryALTinterwordspacing
\emph{Implementation agreement 400{ZR}}, OIF-400ZR-01.0, Optical
  Internetworking Forum, Mar. 2020. [Online]. Available:
  \url{https://www.oiforum.com/wp-content/uploads/OIF-400ZR-01.0_reduced2.pdf}
\BIBentrySTDinterwordspacing

\bibitem{bajaj2022deep}
V.~Bajaj, F.~Buchali, M.~Chagnon, S.~Wahls, and V.~Aref, ``Deep neural
  network-based digital pre-distortion for high baudrate optical coherent
  transmission,'' \emph{J. of Lightw. Technol.}, vol.~40, no.~3, pp. 597--606,
  2022.

\bibitem{wu2021low}
Y.~Wu, U.~Gustavsson, A.~Graell~i Amat, and H.~Wymeersch, ``Low complexity
  joint impairment mitigation of {I/Q} modulator and {PA} using neural
  networks,'' \emph{IEEE J. Sel.Areas Commun.}, vol.~40, no.~1, pp. 54--64,
  2021.

\bibitem{8555649}
X.~Cheng, D.~Liu, Z.~Zhu, W.~Shi, and Y.~Li, ``A {R}es{N}et-{DNN} based channel
  estimation and equalization scheme in {FBMC}/{OQAM} systems,'' in \emph{Proc.
  IEEE Int. Conf.on Wireless Commun. Signal Proc. (WCSP)}, 2018.

\bibitem{morgan2006generalized}
D.~R. Morgan, Z.~Ma, J.~Kim, M.~G. Zierdt, and J.~Pastalan, ``A generalized
  memory polynomial model for digital predistortion of {RF} power amplifiers,''
  \emph{IEEE Trans. Signal Proc.}, vol.~54, no.~10, pp. 3852--3860, 2006.

\bibitem{kamalov2018evolution}
V.~Kamalov \emph{et~al.}, ``Evolution from 8{QAM} live traffic to {PS} 64-{QAM}
  with neural-network based nonlinearity compensation on 11000 km open subsea
  cable,'' in \emph{Proc. Opt. Fiber Commun. Conf. (OFC)}, 2018, p. Th4D.5.

\bibitem{diedolo2022nonlinear}
F.~Diedolo, G.~B{\"o}cherer, M.~Sch{\"a}dler, and S.~Calabr{\'o}, ``Nonlinear
  equalization for optical communications based on entropy-regularized mean
  square error,'' in \emph{Proc. Eur. Conf. Opt. Commun. (ECOC)}, 2022, p.
  We2C.2.

\bibitem{kim2001digital}
J.~Kim and K.~Konstantinou, ``Digital predistortion of wideband signals based
  on power amplifier model with memory,'' \emph{Electron. Lett.}, vol.~37,
  no.~23, p.~1, 2001.

\bibitem{liu2018beam}
X.~Liu \emph{et~al.}, ``Beam-oriented digital predistortion for 5{G} massive
  {MIMO} hybrid beamforming transmitters,'' \emph{IEEE Trans. Microw. Theory
  Tech.}, vol.~66, no.~7, pp. 3419--3432, 2018.

\bibitem{PA_model}
Y.~Wu, J.~Song, C.~Häger, U.~Gustavsson, A.~Graell~i Amat, and H.~Wymeersch,
  ``Symbol-based over-the-air digital predistortion using reinforcement
  learning,'' in \emph{Proc. Int. Conf. Commun. (ICC)}, 2022, pp. 2615--2620.

\bibitem{directlearning_DPD}
G.~Paryanti, H.~Faig, L.~Rokach, and D.~Sadot, ``A direct learning approach for
  neural network based pre-distortion for coherent nonlinear optical
  transmitter,'' \emph{J. Lightw. Technol.}, vol.~38, no.~15, pp. 3883--3896,
  2020.

\end{thebibliography}
\end{document}